\documentclass[11pt,a4paper]{article}
\usepackage{jheppub,amscd,enumerate,xcolor}

\newcommand{\p}[1]{(\ref{#1})}

\newcommand{\cN}{{\cal N}}

\newcommand{\cS}{{\cal S}}

\newcommand{\bD}{{\overline D}{}}

\newcommand{\bQ}{{\overline Q}{}}
\newcommand{\bS}{{\overline S}{}}

\newcommand{\bxi}{{\bar\xi}}

\newcommand{\bpsi}{{\bar\psi}{}}

\newcommand{\be}{\begin{equation}}
\newcommand{\ee}{\end{equation}}
\newcommand{\bea}{\begin{eqnarray}}
\newcommand{\eea}{\end{eqnarray}}

\newcommand{\ba}{\begin{array}} \newcommand{\ea}{\end{array}}

\def\im{{\rm i\,}}
\def\sfrac#1#2{{\textstyle\frac{#1}{#2}}}

\newcommand{\nn}{\nonumber}

\title{(super)Schwarzian mechanics }
\author{Nikolay Kozyrev}
\author{and Sergey Krivonos}
\affiliation{Bogoliubov  laboratory of theoretical physics, \\ Joint institute for nuclear research, \\
Joliot-Curie, 6, 141980 Dubna, Russia}
\emailAdd{nkozyrev@theor.jinr.ru}
\emailAdd{krivonos@theor.jinr.ru}
\abstract{In this paper we revisit the construction of supersymmetric Schwarzians using nonlinear realizations.  We show that $\cN=0,1,2,3,4$ supersymmetric Schwarzians can be systematically obtained as certain projections of Maurer-Cartan forms of superconformal groups after imposing simple conditions on them. We also present the supersymmetric Schwarzian actions defined as the integrals of products of Cartan forms.
In contrast with the previous attempts to obtain the super-Schwarzians within
nonlinear realizations technique, our set of constraints does not impose any
restriction on the super-Schwarzians.}
\arxivnumber{2111.04643}

\begin{document}

\maketitle
\flushbottom

\setcounter{equation}{0}
\section{Introduction}
The holographic relationship between the Sachdev-Ye-Kitaev model \cite{SYK1,SYK2}
and Jackiw-Teitelboim gravity \cite{JT1,JT2} gives rise to numerous research
activities to the Schwarzian action (see e.g. \cite{1}), which provides the boundary description of the bulk JT gravity. The Schwarzian derivative
$\left\{ t, \tau\right\}$ defined as
\be\label{SchwDer}
\left\{ t, \tau\right\} = \frac{\dddot t}{\dot t} - \frac{3}{2} \left( \frac{\ddot t}{\dot t}\right)^2 , \;\; \dot t = \partial_\tau t,
\ee
itself appears in seemingly unrelated fields of physics and mathematics (see e.g. \cite{SW1}).

The action of the bosonic Schwarzian mechanics reads (see e.g. \cite{1})
\be\label{action0}
S_{schw}[t] = - \frac{1}{2} \int d \tau \left( \left\{ t, \tau\right\}+
2 m^2 {\dot t}^2\right).
\ee
Remarkably, the equation of motion of this higher-derivative action is equivalent just to
\be\label{eom1}
\frac{d}{d\tau}\left[  \left\{ t, \tau\right\}+ 2 m^2 {\dot t}^2\right] =0 .
\ee
The characteristic feature of the Schwarzian derivative \p{SchwDer} is its invariance under $SL(2,\mathbb{R})$ M\"{o}bius transformations acting on $t[\tau]$ via
\be\label{sl2a}
t \rightarrow \frac{a t +b}{c t + d} .
\ee
The presence of $m^2 {\dot t}{}^2$ term in the action \p{action0} modifies the realization of $SL(2,\mathbb{R})$
symmetry. The simplest way to understand the modification is to notice that the action \p{action0} can
be represented as \cite{1}
\be\label{F}
S_{schw}[t] = - \frac{1}{2} \int d\tau \left\{ F, \tau \right\}, \qquad F[\tau] = \frac{\tan \left( m t[\tau]\right)}{m},
\ee
and, therefore, the action \p{action0} possesses the $SL(2,\mathbb{R})$ invariance via
\be\label{sl2b}
F \rightarrow \frac{a F +b}{c F + d}
\ee
with $F[\tau]$ defined in \p{F}.

Being invariant under $d=1$ conformal transformation, the Schwarzian derivative
naturally appears in the transformations of the  conformal stress tensor $T(z)$ \cite{BPZ}
\be
T(z) = \left( \frac{d {\tilde z}}{d z}\right)^2 {\tilde T}(\tilde z)+\left\{ {\tilde z},z\right\} .
\ee
The $\cN{=}1,2,3,4$ supersymmetric generalization of the Schwarzian derivative
are present in the transformation properties of the current superfield $J^{(\cN)}(Z)$ generating $\cN$ - extended superconformal transformations \cite{schoutens}. Thus, we have complete zoo of the supersymmetric Schwarzians.

The treatment of the supersymmetric Schwarzians as the anomalous terms in the transformations of the currents superfield $J^{(\cN)}(Z)$ \cite{schoutens} leads to the conclusion that the structure of the (super)Schwarzians is completely defined by the conformal symmetry and, therefore, should exist a different, probably purely algebraic, way to define the (super)Schwarzians. The main property of the (super)Schwarzians, which defines their structure, is their invariance with respect to (super)conformal transformations. The suitable way to construct (super)conformal invariants is the method of nonlinear realizations \cite{coset11,coset12,coset21,coset22} equipped with the inverse Higgs phenomenon \cite{ih}.
Such approach demonstrated how the Schwarzians can be obtained via the non-linear realizations approach, was initiated in \cite{AG1} and then it was applied to different superconformal algebras in \cite{AG2, AG3, AG4, AG5}.
Later on, this approach has been extended to the cases of non-relativistic Schwarzians and Carroll algebra \cite{gomis}.

The reason to prefer the non-linear realizations approach to construction of supersymmetric Schwarzians to the approach related to the superconformal transformations is much wider area of its applications.
Indeed, the non-linear realization method works perfectly for any (super)algebra and the set of invariant Cartan forms can be easily obtained. Thus, the main questions in such approach are
\begin{itemize}
	\item What is the role and source of the ``boundary'' time $\tau$ and its
	supersymmetric partners?
	\item What constraints have to be imposed on the Cartan forms? What forms nullified and how to construct the action from the surviving forms?
	\item What additional technique can be used to simplify the calculations?
\end{itemize}
Of course, these questions were partially analyzed and answered in the papers
\cite{AG1, AG2, AG3, AG4, AG5}. However, some important properties and statements were missing. Moreover, the constraints proposed in these papers looks like the results of illuminating guess. The main puzzle is the fact that the constraints were imposed on the fermionic projections of the forms, but not on the forms themselves. Thus, the questions why it is so and what happens with the full Cartan forms after imposing of such constraints have been not fully analyzed. Finally, in the cases of more complicated superconformal groups the calculations quickly become a rather cumbersome and the standard technique does not help.

In this paper we try to answer these questions. Firstly, we introduce the
``boundary super-space'', where supersymmetry is realized on even and odd coordinates in standard way. Secondly, the constraints will be imposed on the full Cartan forms by either nullifying them or identifying with the ``boundary'' forms. Finally, in the complicated situations we will invoke into game the Maurer-Cartan equations and will demonstrate their usefulness. In particular, we will show that there is only one invariant, $\cN{=}3$ super Schwarzian, in the case of $\cN{=}3$ superconformal symmetry.

\setcounter{equation}{0}
\section{Three steps towards Schwarzian. $\cN{=}0$ case }
In this Section we will repeat the construction of bosonic $\cN{=}0$ Schwarzian
within nonlinear realization approach \cite{AG1}. Mainly, while repeating the steps discussed in \cite{AG1}, we will point the reader's attention at the differences between our approach and those one presented in \cite{AG1}.
\subsection{Step one: the $sl(2,\mathbb{R})$ algebra}
The bosonic conformal group in $d=1$ is infinite-dimensional. Its finite dimensional
$sl(2,\mathbb{R})$ subalgebra spanned by the Hermitian generators of translation $P$, dilatation $D$ and conformal boost $K$, can be fixed by the following relations
\be\label{sl2}
\im \left[ D, P \right] = P,\quad \im \left[ D, K \right] = -K, \quad
\im \left[ K, P \right] = 2 D .
\ee
If we parameterized the $SL(2,\mathbb{R})$ - group element $g$ as
\be\label{param0}
g = e^{\im t \left( P+m^2 K\right)} e^{\im z K} e^{\im u D} ,
\ee
then the Cartan forms
\be\label{CF0}
g^{-1} d g = \im \omega_P P + \im \omega_D D +\im \omega_K K
\ee
would read \cite{IKLe1}
\be\label{forms0}
\omega_P = e^{-u} dt, \quad \omega_D = du - 2 z dt, \quad \omega_K= e^u\left( dz + z^2 dt +m^2 dt\right).
\ee
The infinitesimal  $sl(2,\mathbb{R})$ transformations
\be
g \; \rightarrow \;  g' =  e^{\im a P}\; e^{\im b D}\; e^{\im c K}\, g
\ee
leaving the forms \p{forms0} invariant read
\bea\label{tr0}
\delta t &=& a\, \frac{1+\cos(2 m t)}{2} +b\, \frac{\sin(2 m t)}{2 m}+ c\, \frac{1- \cos(2 m t)}{2 m^2}, \nn \\ \delta u&=& \frac{d}{dt} \delta t, \quad \delta z = \frac{1}{2}  \frac{d}{dt} \delta u - \frac{d}{dt} \delta t\, z .
\eea
At this step our consideration differs from those in \cite{AG1} only by the presence of the term $m^2 K$ in the group element \p{param0}. This additional term generates
the $m$-dependent terms in the Cartan forms \p{forms0} and in the transformations \p{tr0}. As we already discussed in the Introduction this modification is not important.
\subsection{Step two: invariant inverse Higgs conditions}
All Cartan forms in \p{forms0} are invariant with respect to $sl(2,\mathbb{R})$ transformations \p{tr0}. Notice, within the nonlinear realization approach
we implicitly mean that the ``coordinates'' $u$ and $z$ are functions depending on time $t$. However, neither ``time'' $t$, nether its differentials $dt$ are invariant under $sl(2,\mathbb{R})$ transformations \p{tr0}. Thus, to get the invariants one has to introduce the ``boundary time'' $\tau$   {}\footnote{The analogue goes to JT gravity in which
	$\tau$ is the time along the boundary (see e.g. \cite{gomis}).}
and parameterize the form $\omega_P$ as
\be\label{tau1}
\omega_P = e^{-u} dt = d\tau \quad \Rightarrow \quad  {\dot t}  = e^u .
\ee
Let us stress again that the  $\tau$ is a new ``boundary time'' which completely inert under $sl(2,\mathbb{R})$ transformations. Correspondingly, the rest $sl(2,\mathbb{R})$ forms now read
\be\label{formstau}
\omega_D =  \left(\dot{u} - 2 e^u z \right)d\tau ,\quad \omega_K = e^u\left(\dot{z} +e^u \left(z^2+m^2\right)\right) d\tau.
\ee

Now, nullifying the form $\omega_D$ we will express the field $z(\tau)$ in terms of dilaton $u(\tau)$ and then, using \p{tau1}, in terms of ``old time'' $t$:
\be\label{eq1}
\omega_D =0 \quad \Rightarrow \quad  z =\frac{1}{2} e^{-u} \dot{u} = \frac{\ddot t}{2 {\dot t}{}^2} .
\ee
This is particular case of the Inverse Higgs phenomenon \cite{ih}.

\subsection{Step three: the Schwarzian action}
After the Second step only one field, ``old time'' $t(\tau)$, and only one invariant,
form $\omega_K$, remain. Form $\omega_K$ now reads\footnote{The form $\omega_P = d\tau$ is also invariant. However,
	adding this form to the action evidently does not produce new equations of motion.}
\be\label{omegaK}
\omega_K = \frac{1}{2} \left[ \ddot u - \frac{1}{2} {\dot u}{}^2 + 2 m^2 e^{2 u} \right] d\tau =
\frac{1}{2} \left[ \frac{\dddot t}{\dot{t}} -\frac{3}{2} \left( \frac{\ddot t}{\dot t}\right)^2 + 2 m^2 {\dot t}{}^2 \right] d\tau
\ee
Thus, the Schwarzian action \p{action0} can be re-obtained within our approach as
\be\label{action00}
S[t] = - \int \omega_K .
\ee

It proves useful to rewrite the form $\omega_K$ and, therefore, the Schwarzian action
\p{action00} in terms of dilaton $u(t)$ and ``old time'' variable $t$
\be\label{action000}
S[u] = - \int \omega_K = \int dt \left(  \left( \frac{d y}{d t}\right)^2 -
m^2 y^2\right), \qquad y(t) = e^{\frac{1}{2} u(t)}.
\ee
Thus, formally speaking, the action of Schwarzian mechanics is just the action of one dimensional harmonic oscillator rewritten in terms of time variable $t$ depending on new
inert time variable $\tau$.

Note, one may always change the variables $t,z,u$ to the new ones
${\tilde t}, {\tilde z}, \tilde{ u}$ by passing from the  parametrization \p{param0} to $m=0$ one
\be\label{killm}
e^{\im t \left( P+m^2 K\right)} e^{\im z K} e^{\im u D} =
e^{\im \tilde{t} P} e^{\im {\tilde z} K} e^{\im {\tilde u} D}
\ee
It is easy to check that these two parametrizations related as
\be\label{killm1}
m \, {\tilde t} = \tan (m\, t) .
\ee
Clearly, this is just the transformation \p{F}.

Until now our consideration, being purely bosonic one, coincided with those presented in \cite{AG1}.
However, the  generalization to the supersymmetric case will contain some new features. To obtain supersymmetric Schwarzians, one has to consider the proper superalgebra, which differs from \p{sl2} by the presence of supercharges $Q^i$, superconformal charges $S^i$ and, possibly, internal symmetry generators $J^{ij}$. The new ingredients include the introduction of the superconformally inert ``boundary'' superspace  with coordinates $\tau$ and $\theta^i$ using the relations
\be\label{susyidea}
\omega_P = \triangle \tau, \;\; \big( \omega_Q\big)^i = d\theta^i,
\ee
where the forms $\triangle \tau$ and $d\theta^i$ are invariant with respect to standard superspace transformations $\delta\tau \sim \epsilon \theta$, $\delta\theta \sim \epsilon$.

The crucial property of the conditions \p{susyidea} is that they include the Cartan forms $\omega_P$ and $\omega_Q^i$ themselves. Therefore, their invariance under superconformal transformations is manifest.
After imposing condition $\omega_D=0$ also, one should obtain that the remaining forms are composed of supersymmetric Schwarzians and their derivatives.

Note that the forms at the right hand side of equations \p{susyidea} are not arbitrary but constrained by the Maurer-Cartan equations of the respective superconformal algebra. If $\triangle\tau$ is assumed to be generalization of $d\tau$, these equations imply that the forms at the right hand side of equations \p{susyidea} should be standard invariant forms on superspace. This question is studied in detail in Appendices A, B and C.

\setcounter{equation}{0}
\section{ $\cN{=}1$ Schwarzian}
The $\cN{=}1$ super-Schwarzian was firstly introduced in \cite{N1Schw}. Then it has been reproduced in \cite{schoutens}. Within the nonlinear realization of the supergroup $OSp(1|2)$ it was re-constructed in \cite{AG2}. Thus, in this purely illustrative Section, we will show that the our constraints
\p{susyidea} work perfectly, resulting in the proper  $\cN{=}1$ super-Schwarzian.
The crucial property of our construction is that nullifying of the
$\cN{=}1$ super-Schwarzian, similarly to the purely bosonic case, implies
nullifying of all Cartan forms, besides $\omega_P$ and $\omega_Q^i$.
In addition, we will find the $m^2$-modification of the $\cN{=}1$ super-Schwarzian.

The superconformal algebra $osp(1|2)$ contains, in addition to the generators $D$, $P$, $K$, Hermitian fermionic supercharge $Q$ and superconformal charge $S$. Their (anti)commutation relations are:
\bea\label{N1SCA}
&& \im \left[ D, P \right]= P, \quad  \im \left[ D, K \right]= - K, \quad  \im \left[ K, P\right]=2 D,
\nonumber\\
&& \left\{ Q, Q \right\}=2 P, \quad \left\{ S, S \right\}=2 K, \quad  \left\{ Q, S \right\}= - 2 D,
\nonumber\\
&& \im \left[ D, Q \right] = \frac{1}{2} Q,\; \im \left[D, S \right] = -\frac{1}{2} S,\quad \im \left[ K,Q \right] = - S, \;
\im \left[P, S \right]= Q.
\eea
The general element of the $\cN{=}1$ superconformal group $OSp(1|2)$ can be parameterized in analogy with the bosonic case:
\be\label{N1g}
g=e^{\im t\left( P + m^2 K\right)  }\, e^{\xi Q}\, e^{ \psi S} e^{\im z K} e^{\im u D}.
\ee
The Cartan forms
\be\label{N1cfdef}
g^{-1}d g = \im \omega_P P + \omega_Q Q + \im \omega_D D+ \omega_S S+ \im \omega_K K
\ee
explicitly read
\bea\label{N1cf}
&& \omega_P = e^{-u} \triangle t = e^{-u}\left( dt + \im d\xi\,\xi \right), \quad
 \omega_D = d u - 2 z\, \triangle t - 2 \im d\xi\, \psi , \nn \\
&& \omega_K = e^{u}\big( d z +z^2 \triangle t +\im  d \psi\,\psi +2 \im z\, d\xi\, \psi +m^2 (1 - 2 \im \xi \psi) dt\big), \nn \\
&& \omega_Q = e^{-\frac{u}{2}} \left( d\xi +\triangle t\, \psi\right),\quad
\omega_S =e^{\frac{u}{2}}\big(d \psi + z \big( d\xi +\triangle t\, \psi\big)- m^2 \xi dt\big).
\eea
The parameters of the group element \p{N1cfdef} are assumed to be fields that depend on $\tau$ and $\theta$, the coordinates of the $\cN{=}1$ superspace. Supersymmetry is realized on these coordinates in standard way:
\be\label{N1tautheta}
\delta \tau = \im \epsilon\theta, \;\; \delta\theta = \epsilon \;\; \Rightarrow \;\;  \delta d\theta =0, \;\; \delta\triangle \tau =0,\;\;  \triangle \tau =d\tau +\im d \theta\,\theta.
\ee
The inert covariant derivatives defined with respect to $\triangle \tau$ and $d\theta$ \footnote{ These forms and transformation laws, if needed, can be obtained using the coset space techniques, for example, considering coset $\tilde{g} = e^{\im\, \tau P} e^{\theta Q}$, with $P$ and $Q$ forming $\cN{=}1$ Poincare superalgebra. However, they are simple and standard enough to be treated even without reference to nonlinear realizations techniques.}, have the form
\be
d\tau \partial_\tau + d\theta \partial_\theta = \triangle \tau D_\tau + d\theta D \Rightarrow
\left\{ \begin{array}{l}
	D_\tau = \partial_\tau,\\
	D = \frac{\partial}{\partial \theta} -  \im \theta \frac{\partial}{\partial \tau},
\end{array}
\right. \quad D^2 = -\im \partial_\tau.
\ee
Treating all the group parameters in \p{N1g} as fields that depend on $\tau$, $\theta$, we impose the following conditions on the forms $\omega_P, \omega_Q$:
\bea
e^{-u} \triangle t = e^{-u} \left( dt - \im \xi d \xi \right) = \triangle \tau & \Rightarrow &
\left\{ \begin{array}{l}
	{\dot t} - \im \xi \dot \xi = e^u ,\\
	D t+ \im \xi D \xi =0 ,
\end{array} \right. \label{eq11} \\
e^{-\frac{u}{2}} \left( d\xi+ \triangle t \psi \right) = d\theta & \Rightarrow &
\left\{ \begin{array}{l}
	D \xi = e^{\frac{u}{2}}, \\
	\dot\xi + e^u \psi =0 .
\end{array} \right. \label{eq12}
\eea

Finally, one has to nullify the form $\omega_D$ to express the superfield $z$ in terms of $t(\tau,\theta)$ and $\xi(\tau,\theta)$
\be\label{ihN1}
\omega_D= du - 2 z e^u \triangle \tau - 2\im e^{\frac{u}{2}} d\theta \psi =0 \quad \Rightarrow \quad \left\{ \begin{array}{l}
	\dot{u} = 2 z e^u \quad \rightarrow \quad z =\frac{1}{2} e^{-u} {\dot u} ,\\
	D u = 2 \im e^{\frac{u}{2}} \psi .
\end{array} \right.
\ee
Note, only first equation in \p{ihN1} is new, the second one just follows from the equations \p{eq12}. Altogether, equations \p{eq11},\p{eq12}, \p{ihN1} allow to express superfields $u$, $z$, $\psi$  in terms of the Goldstone superfield $\xi(\tau, \theta)$:
\be\label{N1uzpsixi}
u =2 \log( D\xi),\quad \psi =-\frac{\dot\xi}{(D\xi)^2},\quad z =\frac{D{\dot\xi}}{(D\xi)^3}.
\ee
Calculating the remaining Cartan forms $\omega_S$ and $\omega_K$ with relations \p{N1uzpsixi} and their consequences taken into account, one finds that
\bea\label{eqs1a}
&& \omega_S = - \triangle \tau \left[ \frac{\ddot\xi}{D\xi}-2 \frac{{\dot\xi}D{\dot\xi}}{(D\xi)^2}+m^2 (D\xi)^3 \xi\right] \equiv - \triangle \tau \cS(\xi,\{\tau,\theta\}), \nn \\
&& \omega_K = -\triangle \tau D\left(\cS(\xi,\{\tau,\theta\})\right) -\im d\theta \cS(\xi,\{\tau,\theta\}).
\eea
Now, all our forms are expressed in terms of $\cN{=}1$ Schwarzian \cite{N1Schw}
\be\label{Schw1}
\cS_{\cN{=}1}=\cS(\xi,\{\tau,\theta\}) =  \frac{\ddot\xi}{D\xi}-2 \frac{{\dot\xi}D{\dot\xi}}{(D\xi)^2} +m^2 (D\xi)^3 \xi .
\ee
If we compare these expressions \p{Schw1} with those ones from the paper
\cite{AG2}, we conclude that the constant parameters in \cite{AG2} should be chosen as $g=1,p=0$. Thus, in the case of $\cN{=1}$ super-Schwarzian our constraints \p{susyidea} are equivalent (modulo unessential ``$p$''-freedom) to the
constraints introduced in the paper \cite{AG2}. However, already in the case of $\cN{=}2$ super-Schwarzian we will consider in the next Section the preference of our constraints \p{susyidea} becomes evident.

It is clear that the simplest invariant superfield action can be constructed as
\be\label{actionN1}
S_{N1schw} =-\int  \omega_K \wedge \omega_P =-\int \omega_S \wedge \omega_Q =
 \int d\tau d\theta \cS_{\cN{=}1}.
\ee
In component form, it reads
\bea\label{actionN1comp}
S_{N1schw}=-\frac{1}{2} \int d\tau \left[ \frac{\dddot t }{\dot t }  -\frac{3}{2}  \frac{ {\ddot t}\,{}^2 }{ {\dot t}{}^2 } +2m^2 {\dot t}^2 + \im \frac{\dddot\xi \xi +3 \ddot\xi \dot\xi}{\dot t} - \right.\nn \\ \left.- \im \frac{\dddot t \dot\xi \xi + 3 \ddot t \ddot \xi \xi}{\dot t{}^2 } +3\im \frac{{\ddot t}\, {}^2 \dot\xi \xi}{\dot t{}^3}+ 2\im m^2 {\dot t} \xi\dot\xi\right],
\eea
where $t$ and $\xi$ are the first components of respective superfields.

One should stress that in contrast with the bosonic case, the $\cN{=}1$ Schwarzian
\p{Schw1} can not be expressed in terms of the super-dilaton $u$ only due to the presence of the last term which explicitly depends on the fermionic superfield $\xi$, without derivatives.

The infinitesimal $Q$ and $S$ transformations, generated by the element $e^{\epsilon Q + \varepsilon S}$, read
\be\label{osp12}
\delta t =\im \cos(m t) \epsilon \xi - \im \frac{\sin(m t)}{m} \varepsilon \xi, \quad \delta \xi = \epsilon \cos(m t) -\frac{\sin(m t)}{m} \varepsilon .
\ee
All these expressions \p{eqs1a}, \p{Schw1}, \p{actionN1} are invariant with respect to these transformations and, therefore, they are invariant with respect to all $OSp(1|2)$ transformations.

It is completely clear now that if we nullify the super-Schwarzian then the all Cartan forms in \p{N1cfdef} will be equal to zero, besides the forms $\omega_P$ and $\omega_Q$ which will coincide with the ``boundary'' forms $\triangle \tau$ and $d\theta$ \p{N1tautheta}, as expected.

\setcounter{equation}{0}
\section{ $\cN{=}2$ Schwarzian}
The $\cN{=}2$ super-Schwarzian has been introduced in \cite{N2Schw} and then it was re-obtained in \cite{schoutens}. The treatment of the $\cN{=}2$ super-Schwarzian within the nonlinear realization of the $su(1,1|1)$ supergroup
was initiated in \cite{AG2}. The consideration performed in \cite{AG2}
correctly reproduced  $\cN{=}2$ super-Schwarzian but unfortunately the constraints used there imposed the further constraint on the super-Schwarzian
to be a constant. In this Section we will demonstrate that our variant of the constraints \p{susyidea} correctly reproduce $\cN{=}2$ super-Schwarzian,
expressed all  $su(1,1|1)$ Cartan forms  in terms of this super-Schwarzian and its derivatives. Finally, we will show that imposing the constraints on the full Cartan forms makes possible to utilize the Maurer-Cartan equations which drastically simplify all calculations.

In the case of $\cN{=}2$ supersymmetry we are dealing with the $\cN{=}2$
superconformal algebra $su(1,1|1)$ defined by the following relations
\bea\label{N2sca}
&& \im \left[ D, P \right]= P, \quad  \im \left[ D, K \right]= - K, \quad  \im \left[ K, P\right]=2 D,
\nonumber\\
&& \left\{ Q, \bQ \right\}=2 P, \quad \left\{ S, \bS \right\}=2 K, \quad
\left\{ Q, \bS \right\}=  - 2 D + 2 J, \left\{ \bQ, S \right\}= - 2 D - 2 J,
\nonumber\\
&& \im \left[ J ,Q \right] =\frac{1}{2} Q, \; \im \left[ J ,\bQ \right] = -\frac{1}{2} \bQ, \quad
\im \left[ J ,S \right] =\frac{1}{2} S, \; \im \left[ J ,\bS \right] = -\frac{1}{2} \bS, \nn \\
&& \im \left[ D, Q \right] = \frac{1}{2} Q,\; \im \left[ D, \bQ \right] = \frac{1}{2} \bQ,\quad
\im \left[D, S \right] = -\frac{1}{2} S,\;\im \left[D, \bS \right] = -\frac{1}{2} \bS, \nn \\
&& \im \left[ K,Q \right] = - S, \; \im \left[ K,\bQ \right] = - \bS, \quad
\im \left[P, S \right]=  Q, \;\im \left[P, \bS \right]=  \bQ .
\eea

Let us remind the conjugation properties of the generators
\bea\label{conjN2}
\left(P\right)^\dagger, \left(D\right)^\dagger, \left(K\right)^\dagger , \left(J \right)^\dagger& = & P, D, K, -J \nn \\
\left(Q \right)^\dagger, \left(S \right)^\dagger & = & \bQ , \bS.
\eea
The superfields are assumed to depend on the coordinates of the ``boundary'' $\cN{=}2$ superspace $\tau$, $\theta$, $\bar\theta$. The supersymmetry transformations and differential forms, invariant with respect to them, \footnote{If needed, they can be obtained by considering coset $\tilde{g}=e^{\im \tau P}\, e^{\theta Q +\bar\theta \bQ}$} are
\bea\label{N2g0}
\delta \tau = \im \big(  \epsilon\bar\theta + \bar\epsilon\theta \big), \;\; \delta\theta = \epsilon, \;\; \delta\bar\theta = \bar\epsilon, \nn \\
\delta d\theta = 0, \;\; \delta d\bar\theta = 0, \;\; \delta\triangle\tau = 0, \;\; \triangle\tau = d\tau + \im\big(d\bar\theta\, \theta +d\theta\, \bar\theta \big).
\eea
Using the invariant forms \p{N2g0}, one may easily construct the covariant derivatives
\be\label{N2der}
D_\tau = \partial_\tau, \; D = \frac{\partial}{\partial \theta} -\im \bar\theta \frac{\partial}{\partial \tau},\;
\bD = \frac{\partial}{\partial \bar\theta} -\im \theta \frac{\partial}{\partial \tau},  \quad \quad \left\{ D, \bD\right\} = - 2 \im \partial_\tau.
\ee

Similarly to the previously considered cases, we choose the following parametrization of the general element of the $\cN{=}2$ superconformal group $SU(1,1|1)$
\be\label{N2g}
g=e^{\im t (P+m^2 K)}\, e^{\xi Q +\bxi \bQ}\, e^{ \psi S+\bpsi \bS} e^{\im z K} e^{\im u D} e^{ \phi J}
\ee
where the parameters $t, \xi, \bxi, \psi, \bpsi, z, u$ and $\phi$ are, as we stated above, the superfunctions depending on $\{\tau, \theta, \bar\theta\}$. The Cartan forms
\be\label{N2cfdef}
g^{-1}d g = \im \omega_P P+  \omega_Q Q + {\bar\omega}_Q \bQ +\im \omega_D D+  \omega_J J + \omega_S S + {\bar\omega}_S \bS +\im \omega_K K
\ee
explicitly read
\bea\label{N2cf}
&& \omega_P \equiv e^{-u} \triangle t = e^{-u}\left( dt - \im (\xi d\bxi + \bxi d\xi) \right), \nn \\
&& \omega_Q = e^{-\frac{u}{2}+\im \frac{\phi}{2}}  \left( d\xi +\psi\, \triangle t\right),\;
\bar{\omega}_Q = e^{-\frac{u}{2}-\im \frac{\phi}{2}} \left( d\bxi +\bpsi \,\triangle t\right),\nn \\
&& \omega_D = d u - 2 z\, \triangle t - 2 \im (d\xi\bpsi+d\bxi \psi) , \;
\omega_J = d \phi - 2 \psi \bpsi\, \triangle t +  2 (d\bxi\psi-d\xi \bpsi)- 2 m^2 \xi \bxi dt \nn \\
&& \omega_S = e^{\frac{u}{2}+\im \frac{\phi}{2}} \left(d \psi -\im \psi \bpsi d\xi + z \left( d\xi +\psi\, \triangle t\right)\, - m^2 \left(1-\im \bxi \psi \right) \xi dt\right),  \\
&&\bar{\omega}_S  =  e^{\frac{u}{2}-i \frac{\phi}{2}} \left(d \bpsi +\im \psi \bpsi d\bxi + z \left( d\bxi +\bpsi \,\triangle t\right)\,- m^2 \left(1-\im \xi \bpsi \right) \bxi dt\right),\nn \\
&& \omega_K = e^{u}\left( d z +z^2 \triangle t -\im (\psi\, d \bpsi +\bpsi\, d\psi) +2 \im z\, (d\xi\, \bpsi + d\bxi \psi) +m^2 \left( 1+ \im \left( \psi \bxi+\bpsi \xi\right) \right)^2\, dt\right).\nn
\eea

Now, imposing the constraints \p{susyidea}, i.e. identifying the forms $\omega_P, \omega_Q, {\bar \omega}_Q$ \p{N2cf} with $\triangle\tau$, $d\theta$, $d\bar\theta$ \p{N2g0} respectively, will result in the following equations
\bea
e^{-u} \triangle t = e^{-u}\left( dt + \im \left( d \bxi \xi+ d \xi \bxi\right)\right) = \triangle \tau \quad & \Rightarrow &  \quad
\left\{ \begin{array}{l}
	\dot{t}+ \im \left( \dot{\bxi}\xi+\dot{\xi}\bxi\right) = e^u, \\
	D t + \im D\xi\, \bxi =0, \\
	\bD t+ \im \bD \bxi\, \xi =0 ,
\end{array} \right. \label{eqN2t} \\
e^{-\frac{1}{2}(u-\im\, \phi)}\left( d\xi +\psi \triangle t\right) = d\theta  \quad & \Rightarrow &\quad
\left\{ \begin{array}{l}
	\dot{\xi}+ e^u \psi =0, \\
	D \xi =  e^{\frac{1}{2}(u-\im\, \phi)}, \\
	\bD \xi =0 ,
\end{array} \right. \label{eqN2xi} \\
e^{-\frac{1}{2}(u +\im\, \phi)}\left( d\bxi +\bpsi \triangle t\right) = d\bar\theta  \quad & \Rightarrow &\quad
\left\{ \begin{array}{l}
	\dot{\bxi}+ e^u \bpsi =0, \\
	\bD \bxi =  e^{\frac{1}{2}(u + \im\, \phi)}, \\
	D \bxi =0 .
\end{array} \right. \label{eqN2bxi}
\eea
Finally, one has to nullify the form $\omega_D$ :
\be\label{eqN2D}
\omega_D = d u - 2 e^u z\, \triangle \tau  - 2 \im (e^{\frac{1}{2}(u -\im\, \phi)}d\theta\bpsi+e^{\frac{1}{2}(u +\im\, \phi)}d\bar\theta \psi) =0 \quad \Rightarrow \quad
\left\{ \begin{array}{l}
	\dot{u} - 2 e^u z =0, \\
	D u = 2 \im \, e^{\frac{1}{2}(u - \im\, \phi)} \bpsi, \\
	\bD u = 2 \im \, e^{\frac{1}{2}(u + \im\, \phi)} \psi.
\end{array} \right.
\ee
{}From these relations one may obtain several important consequences. In particular, we have
\bea
&& D u = \im D \phi, \bD u =- \im \bD \phi,\quad \Rightarrow \quad \left[D, \bD\right] u = - 2 \dot\phi, \;
\left[ D, \bD\right] \phi = 2 {\dot u}, \label{con1} \\
&& D \bpsi=0, \quad \bD \psi =0,\quad \psi = - \frac{\dot\xi}{D\xi \bD\bxi}, \; \bpsi = - \frac{\dot\bxi}{D\xi \bD\bxi}, \label{con2} \\
&& D \xi \, \bD \bxi = e^u,\;
\dot{u} = \frac{D\dot\xi}{D\xi}+\frac{\bD\dot{\bxi}}{\bD\bxi}, \quad\frac{\bD \bxi}{D \xi} = e^{\im \phi}, \quad \dot\phi = \im\left( \frac{D\dot\xi}{D\xi}-
\frac{\bD\dot{\bxi}}{\bD\bxi}\right) . \label{con3}
\eea
Now, one may check that the form $\omega_J$ reads
\be\label{N2Schw}
\omega_J =
\im\left[ \frac{D\dot{\xi}}{D\xi} -\frac{\bD \dot{\bxi}}{\bD\bxi}-2 \im \frac{\dot\xi \dot{\bxi}}{D\xi \bD\bxi} + 2 \im m^2 \xi \bxi D\xi \bD\bxi\right]\triangle \tau\equiv
\im \, \triangle\tau\, \cS_{\cN{=}2} .
\ee
Thus we see, that $\cN{=}2$ Schwarzian $\cS_{\cN{=}2}$ appears automatically.
One may check that the other Cartan forms, $\omega_S, {\bar\omega}_S$ and  $\omega_K$ can be also expressed in terms of  the $\cN{=}2$ Schwarzian only
\bea
\omega_P &=& \triangle\tau, \omega_Q =d\theta,\; {\bar\omega}_Q = d\bar\theta, \quad
\omega_J = \im \cS_{\cN{=}2} \triangle\tau, \nn \\
\omega_S & = & -\frac{1}{2} \cS_{\cN{=}2}\, d\theta-\frac{\im}{2} \bD\cS_{\cN{=}2} \triangle \tau,\quad
{\bar\omega}_S  =  \frac{1}{2} \cS_{\cN{=}2}\, d{\bar\theta}+\frac{\im}{2}  D \cS_{\cN{=}2} \triangle \tau,\label{finformsN2} \\
\omega_K & =& \frac{1}{2} D \cS_{\cN{=}2} d\theta - \frac{1}{2} \bD \cS_{\cN{=}2} d\bar\theta +\frac{1}{4} \left( \im \left[ D, \bD\right] \cS_{\cN{=}2} - \cS_{\cN{=}2}^2 \right) \triangle \tau. \nn
\eea

The transformation laws of the basic superfields $t$, $\xi$, $\bxi$,  are induced by left multiplication $g^\prime = g_0 g$. In the case of superconformal transformations  $g_0 =e^{\epsilon Q + \bar\epsilon\bQ}e^{\varepsilon S + \bar\varepsilon \bS}$  the transformation laws of $t$ and $\xi$, $\bxi$ read
\bea\label{N2scftr}
\delta t &=& \im \big( \bar\epsilon\xi +\epsilon\bxi  \big)\cos(mt) - \im \frac{\sin(mt)}{m} \big( \bar\varepsilon\xi +\varepsilon\bxi  \big), \nn \\
\delta \xi &=& \cos(mt)\epsilon + \im \epsilon m \sin(mt)\xi\bxi - \frac{\sin(mt)}{m}\varepsilon + \im \varepsilon \cos(mt)\xi\bxi, \\
\delta \bxi &=& \cos(mt)\bar\epsilon - \im \bar\epsilon m \sin(mt)\xi\bxi - \frac{\sin(mt)}{m}\bar\varepsilon - \im \bar\varepsilon \cos(mt)\xi\bxi.\nn
\eea
The modified $\cN{=}2$ Schwarzian  $\cS_{\cN{=}2}$
\be\label{N2Schwm}
\cS_{\cN{=}2}  = \frac{D\dot{\xi}}{D\xi} -\frac{\bD \dot{\bxi}}{\bD\bxi}-2 \im \frac{\dot\xi \dot{\bxi}}{D\xi \bD\bxi} + 2 \im m^2 \xi \bxi D\xi \bD\bxi
\ee
is invariant with respect to these transformations.
 Thus one can expect that the proper superfield Schwarzian action reads
\bea\label{N2schwact}
S_{N2schw} &=& -\frac{\im}{2} \int d\tau \, d\theta\,d\bar\theta \cS = -\frac{1}{2}\int \omega_J\, \wedge \,\omega_Q \,\wedge \,{\bar\omega}_Q =\nn \\
&=& \im \int \omega_P\, \wedge\, \omega_S\, \wedge\, {\bar\omega}_Q  =-\im \int \omega_P\, \wedge\, \omega_Q\, \wedge\, {\bar\omega}_S.
\eea
Evaluating the integral, one can find the component action
\bea\label{N2schwcompact}
S_{N2schw} = -\frac{1}{2}\int d\tau  \left[ \frac{\dddot t}{\dot t} - \frac{3}{2} \frac{{\ddot t}\, {}^2}{{\dot t} {}^2}  + 2m^2 \dot t{}^2 - \frac{1}{2}{\dot\phi}^2- \im \frac{\dddot t \big( \dot \xi \bxi + \dot\bxi \xi \big)}{\dot t} - 3\im \frac{\ddot t \big( \ddot\xi \bxi + \ddot\bxi \xi\big)}{{\dot t}{}^2}+\right. \nn \\ \left.+\im \frac{\dddot \xi \bxi + \ddot \bxi \xi + 3 \ddot\xi \dot\bxi +3 \ddot\bxi \dot\xi}{\dot t} + 3\im \frac{\ddot t{}^2 \big( \dot\xi \bxi + \dot\bxi \xi\big)}{\dot t{}^2}- 2\frac{\dddot t \xi\bxi\dot\xi\dot\bxi}{\dot t{}^3} + \frac{\xi\bxi\big( \dddot \xi \dot\bxi - \dddot\bxi \dot \xi  \big)}{\dot t{}^2} -\right. \nn \\ \left. -  3\frac{\dot \xi\dot\bxi \big( \ddot\xi\bxi - \ddot\bxi \xi  \big)}{\dot t{}^2} + 9 \frac{\ddot t{}^2 \xi\bxi\dot\xi\dot\bxi}{\dot t{}^4} +3\frac{\xi\bxi\ddot\xi\ddot\bxi}{\dot t{}^2} - 6\frac{\ddot t \xi\bxi \big( \ddot\xi\dot\bxi - \ddot\bxi\dot\xi \big)}{\dot t{}^3}-2\frac{\dot\phi \dot\xi\, \dot\bxi}{\dot t}-\right. \\ \left. -2\im m^2 \dot t \big( \dot\xi \bxi + \dot\bxi\xi \big)+2m^2 \dot\phi \dot t \xi\bxi
\right]. \nn
\eea
where $t$, $\xi$, $\bxi$ and $\phi$ are the first components of respective superfields.

The calculations leading to the expressions \p{finformsN2} are rather involved.
They become more and more complicated while passing to higher supersymmetries.
However, the fact that our constraints \p{susyidea} are imposed on the Cartan forms opens the way to use the Maurer-Cartan equations which drastically simplify the calculations. The $\cN{=}2$ case provides a nice possibility to demonstrate on the simplest example how everything is  working on. We put
this consideration in the Appendix A.

Comparing our expressions for the final Cartan forms \p{finformsN2} with the constraints which were used in the paper \cite{AG2} we conclude that the
constraint
$$\omega_S|_{d\theta} = -\frac{1}{2} \cS_{\cN{=}2} =p$$
immediately restricts $\cN{=}2$ Schwarzian to be a constant. Clearly, this condition is unreasonably strong. Thus, the $\cN{=}2$ case is the first one in which our set of constraints \p{susyidea} and $\omega_D=0$ becomes preferable
with respect to those ones formulated in \cite{AG2}.

\setcounter{equation}0
\section{ $\cN{=}3$ Schwarzian}
The $\cN{=3}$ super Schwarzian
\be\label{N3Schw}
\cS_{\cN{=}3} = \frac{1}{2} \frac{\epsilon_{pqr}D_p\xi_n\, D_q D_r \xi_n}{D_k \xi_l \, D_k \xi_l}
\ee
has been introduced in \cite{schoutens}. Then it was re-obtained within the
nonlinear realization of the supergroup $OSp(3|2)$ in \cite{AG4}. However,
the constraints introduced in \cite{AG4} lead, besides the $\cN{=3}$ super Schwarzian $\cS_{\cN{=}3}$, to some new  $OSp(3|2)$ invariants with unclear geometric meaning. In this Section we will demonstrate that our constraints \p{susyidea} and $\omega_D=0$ being applied to superalgebra $osp(3|2)$
results in the Cartan forms expressed in terms of $\cS_{\cN{=}3}$ and its derivatives only.

The $osp(3|2)$ superalgebra contains 6 bosonic $(P,D,K,J_i)$ and 6 fermionic generators $Q_i,S_j$ obeying the following (anti)commutators:
\bea\label{N3SCA}
&& \im \left[ D, P \right]= P, \quad  \im \left[ D, K \right]= - K, \quad  \im \left[ K, P\right]=2 D,
\nn \\
&& \left\{ Q_i, Q_j \right\}=2 \delta_{ij} P, \quad \left\{ S_i, S_j \right\}=2 \delta_{ij} K, \quad
\left\{ Q_i, S_j \right\}= - 2 \delta_{ij} D -\epsilon_{ijk} J_k,
\nn \\
&& \im \left[ D, Q_i \right] = \frac{1}{2} Q_i,\; \im \left[D, S_i \right] = -\frac{1}{2} S_i,\quad \im \left[ K,Q_i \right] = - S_i, \;
\im \left[P, S_i \right]= Q_i, \nn \\\
&& \im \left[ J_i, Q_j \right] = \epsilon_{ijk} Q_k, \quad \im \left[ J_i, S_j \right] = \epsilon_{ijk} S_k, \quad \im \left[ J_i, J_j \right] = \epsilon_{ijk} J_k .
\eea
Here, all generators are chosen to be hermitean, $i,j,k\ldots = 1,2,3$ and $\epsilon_{ijk}$ is completely antisymmetric symbol, $\epsilon_{123}=1$. We parameterize the general element of the $\cN{=}3$ superconformal group as\footnote{From now on, to simplify all calculations,  we will omit the term $m^2 K$ in the group element \p{N3g}.}
\be\label{N3g}
g=e^{\im t P}\, e^{ \xi_i Q_i }\, e^{ \psi_j S_j} e^{\im z K} e^{i u D} e^{\im \phi_i  J_i},
\ee
with the invariant Cartan forms defined as
\be\label{N3cfdef}
\Omega = g^{-1}d g = \im \omega_P P+ \left(\omega_Q\right)_i Q_i + \im\omega_D D+
 \im \left(\omega_J\right)_{i} J_{i}   +  \left(\omega_S\right)_i S_i + \im \omega_K K.
\ee
The forms of $P$, $D$, $K$ generators read
\bea\label{N3cfPDK}
\omega_P &=& e^{-u} \left( dt - \im \, \xi_i d\xi_i \right)\equiv  e^{-u} \triangle t, \nn \\
\omega_D &=& d u - 2 z \triangle t - 2 \im \,d \xi_i \psi_i , \\
\omega_K &=&  e^{u} \left( d z +z^2 \triangle t - \im \, \psi_i d\psi_i  - 2 \im \, z \psi_i d \xi_i \right).\nn
\eea
The forms of fermionic generators and $J_i$, unlike \p{N3cfPDK}, include rotations, induced by the exponent $e^{\im \phi_k J_k}$, which can be parameterized with $SO(3)$ matrix $M_{ij}$:
\bea\label{N3cfQSJ}
\big(\omega_Q \big)_i &=& \big({\hat\omega}_Q \big)_j M_{ij}, \;\; \big(\omega_S \big)_i=\big({\hat\omega}_S \big)_j M_{ij}, \;\; \big(\omega_J \big)_i = \big({\hat\omega}_J \big)_j M_{ij} + \frac{1}{2}\epsilon_{ijk} dM_{jm} M_{km}, \nn \\
M_{ij} &=& \big(e^{q}\big)_{ij}, \;\; q_{ij} = \epsilon_{ijk}\phi_k, \;\; \big( M^{-1} \big)_{ij} = M_{ji}, \;\; \det M =1.
\eea
The hatted forms here are
\bea\label{N3cfQSJhat}
\left(\hat\omega_Q\right)_i & = & e^{-\frac{u}{2}} \left( d \xi_i + \triangle t \psi_i\right), \nn \\
\left( \hat\omega_J\right)_{i} & =&-\im \epsilon_{ijk} \left( \psi_j d\xi_k + \frac{1}{2} \triangle t \psi_j \psi_k \right),  \\
\left({\hat\omega}_S\right)_i &=&  e^{\frac{u}{2}} \left( d \psi_i - \im \psi_i \psi_j d\xi_j + z\left( d\xi_i + \triangle t \, \psi_i \right)\right).\nn
\eea
We treat the parameters of the $OSp(3|2)$ group element as superfields that depend on the coordinates of the $\cN{=}3$ superspace, $\tau$ and $\theta_i$. The $\cN{=}3$ supersymmetry is realized on these coordinates in standard way,\footnote{Just as before, these forms and transformation laws can be obtained by considering coset element ${\tilde g}=e^{\im \tau P}\, e^{\theta_i Q_i }$, where $P$ and $Q_i$ form $\cN{=3}$ Poincare superalgebra}
\be\label{N3g0}
\delta \tau = \im \epsilon_i \theta_i, \;\; \delta \theta_i = \epsilon_i, \;\; \delta d\theta =0, \;\; \delta\triangle \tau = 0, \;\; \triangle\tau = d\tau + \im d\theta_i\, \theta_i.
\ee
Correspondingly,  the $\cN{=}3$ covariant derivatives read
\be\label{N3der}
D_\tau = \partial_\tau, \; D_i = \frac{\partial}{\partial \theta_i} -i \theta_i \frac{\partial}{\partial \tau},  \quad \quad \left\{ D_i, D_j\right\} = - 2 i \delta_{ij} \partial_\tau .
\ee

Just as before, we enforce the following invariant constraints on the Cartan forms
\be\label{N3constr}
\omega_P = \triangle \tau, \;\; \big(\omega_Q  \big)_i = d\theta_i, \;\; \omega_D =0.
\ee
As was suggested by the results of the previous Section, much information about the Schwarzian can be obtained by analyzing structure of the Cartan forms with the help of Maurer-Cartan equations
\be
d_2\Omega_1 - d_1 \Omega_2 = \big[ \Omega_1, \Omega_2  \big].\nn
\ee
Leaving detailed description of this calculation to the Appendix B, we present here only the result:
\bea\label{N3formssol}
\omega_P &=&\triangle \tau, \;\; \big(\omega_Q \big)_i = d\theta_i, \;\; \omega_D =0, \;\; \big(\omega_J\big)_i = \im \triangle\tau D_i \cS + d\theta_i\, \cS, \nn \\
\big(\omega_S\big)_i &=& \triangle \tau \Big( \cS D_i \cS - \frac{1}{2}  \epsilon_{ipq} D_p D_q \cS \Big) + \im \epsilon_{ijk}d\theta_j D_k \cS,\\
\omega_K &=&  \triangle \tau \Big(   -\im \cS \dot\cS + \frac{1}{6}\big(\epsilon_{pqr} D_p D_q D_r \cS  \big) - D_k \cS D_k\cS    \Big) + \im d\theta_i \Big(  \cS D_i \cS - \frac{1}{2}  \epsilon_{ipq} D_p D_q \cS    \Big).\nn
\eea
Therefore, all the Cartan forms \p{N3cfdef} can be written in terms of just one fermionic superfield $\cS$ and its derivatives, with no constraints on $\cS$ coming from Maurer-Cartan equations. It is natural to identify this fermionic superfield with the $\cN{=}3$ super-Schwarzian:
\be
\cS = \cS_{\cN{=}3}.
\ee

To relate $\cS_{\cN{=}3}$ to the group superfield parameters, one should study the conditions \p{N3constr} explicitly, writing all their projections with respect to $\triangle \tau$ and $d\theta_i$:
\bea\label{N3irrcond}
\omega_P = \triangle \tau \;\; \Rightarrow \;\; \dot t + \im \dot\xi_i \, \xi_i = e^u, \;\; D_i \, t + \im D_i \xi_j \, \xi_j =0, \nn \\
\big(\omega_Q\big)_i = d\theta_i \;\; \Rightarrow \;\; D_j \xi_k = e^{u/2}  M_{jk}, \;\; \psi_k =- e^{-u}\dot\xi_k.
\eea
The condition $D_i \, t + \im D_i \xi_j \, \xi_j =0$ can be considered as primary one. From it, one can obtain
\bea\label{N3irrcond2}
&&D_i\big( D_j t + \im D_j\xi_k\, \xi_k  \big) + D_j\big( D_i t + \im D_i\xi_k\, \xi_k  \big) =0 \; \Rightarrow \; \nn \\&& -2\im \delta_{ij} \big( \dot t + \im \dot\xi_k \xi_k\big) +2 \im D_i\xi_k \, D_j\xi_k =0 \; \Rightarrow \; D_i \xi_k \, D_j \xi_k = \delta_{ij}e^u,
\eea
and $D_i \xi_j$ is proportional to the orthogonal matrix. This way one can also obtain the derivative of $u$, $D_i e^{u} = -2\im D_i\xi_j \, {\dot\xi}_j$. Condition $\omega_D =0$ then just expresses $z$ in terms of $u$ as $z = \frac{1}{2}e^{-u}\dot u$.

With these conditions taken into account, one can write down $d\theta_p$ projection of $\big( \omega_J  \big)_i$ as
\bea\label{N3schwarzian1}
&&\big(\omega_J \big)_i = \ldots + d\theta_p \Big[  -\im M_{ik} \epsilon_{klm}D_p \xi_l \psi_m + \frac{1}{2}\epsilon_{ijk} e^{u}D_p D_j\xi_m \; D_k \xi_m   \Big] =  \\
&&=\ldots+ d\theta_p \Big[ +\im e^{-3u/2}D_i \xi_k \, D_p \xi_l\, \epsilon_{klm}\dot{\xi}_m  -2\im \epsilon_{ipk}e^{-u}D_k\xi_m\, \dot\xi_m + \frac{1}{2}\epsilon_{ijk} e^{-u} D_p \xi_m \, D_j D_k \xi_m  \Big].\nn
\eea
To proceed further, one should note that, as a consequence of \p{N3irrcond}, the fermionic superfield $\xi_i$ satisfies a quadratic relation
\be\label{N3qferm}
\epsilon_{ipq}D_{m}\xi_n \, D_{p}D_q \xi_n = 2\im \epsilon_{imk} D_k \xi_n\, \dot\xi_n + \frac{1}{3}\delta_{im} \epsilon_{pqr}D_p\xi_n\, D_q D_r \xi_n.
\ee
To obtain it, one should take the relation $D_m \big(\epsilon_{ipq} D_p D_q t  \big) = -2\im \epsilon_{imk}D_k\dot t + \frac{1}{3}\delta_{im} \big(\epsilon_{pqr} D_p D_q D_r t \big)$, which follows just from anticommutation relations of $D_i$, and substitute $D_i t = -\im D_i \xi_j \, \xi_j$ \p{N3irrcond}.
Also taking into account that
\be\label{N3detrel}
D_i \xi_k \, D_p \xi_l\, \epsilon_{klm}\dot{\xi}_m = \det \big( D\xi  \big) \epsilon_{ipn} \big( D\xi^{-1} \big)_{qn}\dot{\xi}_q = e^{3u/2} \epsilon_{ipn}e^{-u}D_n\xi_q \, \dot{\xi}_q,
\ee
the form $\big(\omega_J \big)_i$ \p{N3schwarzian1} reduces to
\be\label{N3schwarzian2}
\big(\omega_J \big)_i = \ldots + d\theta_i \cS_{\cN{=}3}, \;\; \cS_{\cN{=}3} =\frac{1}{6}e^{-u}\epsilon_{pqr}D_p\xi_n\, D_q D_r \xi_n = \frac{1}{2} \frac{\epsilon_{pqr}D_p\xi_n\, D_q D_r \xi_n}{D_k \xi_l \, D_k \xi_l}.
\ee
Obtained $\cS_{\cN{=}3}$ is just the already known $\cN{=}3$ Schwarzian
\cite{schoutens,AG4}.

The obvious candidate for Schwarzian action in $\cN{=}3$ case is
\be\label{N3schwact}
S_{N3schw} = -\frac{1}{6}\int d\tau d\theta_i d\theta_j d\theta_k \epsilon^{ijk} \cS_{\cN{=}3}.
\ee
This is further substantiated by the fact that $d\tau$ projection of $\omega_K$, which defines the component Schwarzian action, contains $\epsilon_{pqr} D_p D_q D_r \cS_{\cN{=}3}$. Using this property to calculate component form of \p{N3schwact}, one obtains
\bea\label{N3compact}
S_{N3schw}&=& -\frac{1}{6} \int d\tau \epsilon_{ijk}D_i D_j D_k \cS_{\cN{=}3} =  -\frac{1}{2}\int d\tau \left[ \frac{\partial_\tau^2 \big( \dot t+ \im \dot\xi_i\, \xi_i \big)}{\dot t+ \im \dot\xi_i\, \xi_i} - \frac{3}{2}\left( \frac{\partial_\tau \big(  \dot t+ \im \dot\xi_i\, \xi_i \big)}{ \dot t+ \im \dot\xi_i\, \xi_i}\right)^2 - \right. \nn \\&& \left. - 2\im \frac{{\dot\xi}_i {\ddot \xi}_i}{\dot t+ \im \dot\xi_i\, \xi_i} +2\im s \dot s -2\im \frac{\dot{M}_{km}M _{kn}{\dot\xi}_m {\dot \xi}_n}{\dot t+ \im \dot\xi_i\, \xi_i}-\dot{M}_{kl} \dot{M}_{kl}\right].
\eea
Here, $t$, $\xi_i$ and $M_{ij}$ are the first components of respective superfields, and $s$ is the first (fermionic) component of the $\cN{=}3$ Schwarzian \p{N3schwarzian2}. It should be taken as independent, as calculating $\big\{ D_i, D_j\}\cS_{\cN{=}3}$ using $D_i\cS_{\cN{=}3}$ extracted from $\triangle\tau$ projection of $\big(\omega_J\big)_i$ Cartan form,
\be\label{N3DS}
D_i\cS_{\cN{=}3} = \frac{1}{2}e^{-u}M_{ik}\epsilon_{klm}{\dot\xi}_l {\dot\xi}_m + \frac{\im}{2}\epsilon_{ijk} M_{jm}\dot{M}_{km},
\ee
one arrives just to an identity.

In terms of Cartan forms, the integral \p{N3schwact} could be written as
\be\label{N3schwactform1}
 S_{N3schw} =-\frac{1}{6} \int \omega_P \, \wedge \, \big(\omega_Q \big)_i \, \wedge \, \big(\omega_Q \big)_j \, \wedge \big(\omega_J \big)_k \, \epsilon^{ijk}.
\ee

\setcounter{equation}0
\section{ $\cN{=}4$ Schwarzian}
Let us, finally, consider the construction of the $\cN{=}4$ Schwarzian.
In this paper, we do not make an attempt to consider the $\cN{=}4$ superconformal algebra $D(2,1,\alpha)$ and concentrate on its particular limit, the $su(1,1|2)$ superalgebra. The corresponding  $\cN{=}4$ Schwarzian has been
constructed in \cite{schoutens,MU} and then it was re-obtained within the nonlinear realization $su(1,1|2)$ superalgebra in \cite{AG3}. In this Section we are going to use our set of constraints \p{susyidea} and $\omega_D=0$ to
demonstrate that all the Cartan forms in this case can be expressed in terms of
$\cN{=}4$ Schwarzian and its derivatives.

The $su(1,1|2)$ superalgebra contains usual conformal generators $P$, $D$, $K$, supersymmetric and superconformal charges $Q_\alpha$, $\bQ{}^\alpha = \big(  Q_\alpha\big)^\dagger$, $S_\alpha$, $\bS{}^\alpha = \big( S_\alpha \big)^\dagger$ and generators of the $su(2)$ subalgebra $T_\alpha{}^\beta = - \big( T_\beta{}^\alpha \big)^\dagger$, $T_\alpha{}^\alpha =0$ :
\bea\label{2Nsuperconf}
\big[ D,P  \big]&=&-\im P, \;\; \big[ D,K  \big]=\im K, \;\; \big[ P,K  \big]=2\im D, \;\; \big\{ Q_\alpha,\bQ^\beta  \big\}=2\delta_\alpha^\beta P, \;\; \big\{ S_\alpha,\bS^\beta  \big\}=2\delta_\alpha^\beta K, \nn \\
\big\{  Q_\alpha, \bS^\beta\big\} &=& -2\delta_\alpha^\beta D -2 T_\alpha{}^\beta, \;\; \big\{  \bQ^\alpha, S_\beta\big\} = -2\delta^\alpha_\beta D +2 T_\beta{}^\alpha, \\
\big[ D, Q_\alpha\big] &=& -\sfrac{\im}{2}Q_\alpha,\;\;  \big[ D, \bQ^\alpha\big] = -\sfrac{\im}{2}\bQ{}^\alpha,\;\; \big[ D, S_\alpha\big] = \sfrac{\im}{2}S_\alpha,\;\;  \big[ D, \bS{}^\alpha\big] = \sfrac{\im}{2}\bS{}^\alpha,\nn \\
\big[ K, Q_\alpha\big] &=&\im S_\alpha,\;\;  \big[ K, \bQ{}^\alpha\big] = \im \bS{}^\alpha,\;\; \big[ P, S_\alpha\big] = -\im Q_\alpha,\;\;  \big[ P, \bS{}^\alpha\big] = -\im\bQ{}^\alpha.\nn
\eea
The generators $D,K,P$ commute with $su(2)$; the commutators of $su(2)$ with themselves and fermionic generators read
\bea\label{2Nsuperconf2}
\big[ T_\alpha{}^\beta, T_\mu{}^\nu  \big] &=& \im \big( \delta_\mu^\beta T_\alpha{}^\nu  - \delta_\alpha^\nu T_\mu{}^\beta  \big), \nn \\
\big[ T_\alpha{}^\beta , Q_\gamma \big] &=& \im \left( \delta_\gamma^\beta Q_\alpha -\frac{1}{2}\delta_\alpha^\beta Q_\gamma  \right), \;\; \big[ T_\alpha{}^\beta ,\bQ^\gamma \big] = -\im \left( \delta_\alpha^\gamma \bQ^\beta - \frac{1}{2}\delta_\alpha^\beta \bQ^\gamma  \right), \nn \\
\big[ T_\alpha{}^\beta , S_\gamma \big] &=& \im \left( \delta_\gamma^\beta S_\alpha - \frac{1}{2}\delta_\alpha^\beta S_\gamma  \right), \;\; \big[ T_\alpha{}^\beta ,\bS^\gamma \big] = -\im \left( \delta_\alpha^\gamma \bS^\beta - \frac{1}{2}\delta_\alpha^\beta \bS^\gamma  \right).
\eea
Here, indices $\alpha,\beta,\ldots=1,2$ can be raised and lowered with help of antisymmetric tensors $\epsilon_{\alpha\beta}$, $\epsilon^{\alpha\beta}$, $\epsilon_{\alpha\beta}\epsilon^{\beta\gamma} = \delta_\alpha^\gamma $, $\epsilon_{12} = \epsilon^{21}=1$.

The $SU(1,1|2)$ group element can be parameterized as
\be\label{N4coset}
g = e^{\im tP}e^{\xi^\alpha Q_\alpha + \bxi_\alpha\bQ^\alpha} e^{\psi^\alpha S_\alpha + \bpsi_\alpha\bS^\alpha}e^{\im z K}  e^{\lambda_\beta{}^\alpha T_\alpha{}^\beta} e^{\im u D}.
\ee
The Cartan forms defined in  standard way
\be\label{N4cfdef}
\Omega = {g}^{-1}d{g} = \im \omega_P P + \im \omega_K K +\im \omega_D D + \big(\omega_Q\big)^\alpha Q_\alpha +\big(\bar\omega_Q\big)_\alpha\bQ^\alpha + \big(\omega_S\big)^\alpha S_\alpha +\big(\bar\omega_S\big)_\alpha\bS^\alpha+\big( \omega_T \big)_\beta{}^\alpha T_\alpha{}^\beta
\ee
are rather involved. Explicitly, the forms of the scalar  bosonic generators, $P,D$ and $K$ read
\bea\label{2Ntildegmformsbos}
\omega_{P} &=& e^{-u}\triangle t, \;\; \triangle t = dt +\im \big( d\xi^\alpha\bxi_\alpha + d\bxi_\alpha\xi^\alpha  \big), \;\; \omega_{D} = du -2\im \big(d\xi^\alpha\bpsi_\alpha + d\bxi_\alpha\psi^\alpha  \big)-2z\triangle t , \nn \\
\omega_{K} &=& e^u \big[ dz + z^2\triangle t + \im \big(d\psi{}^\alpha\bpsi_\alpha + d\bpsi_\alpha\psi{}^\alpha  \big) + \triangle t \big( \psi^\alpha\bpsi_\alpha \big)^2 +2\big(d\xi^\alpha\bpsi_\alpha - d\bxi_\alpha\psi^\alpha  \big)\big( \psi^\beta\bpsi_\beta \big) +\nn \\ &&+2\im z \big(d\xi^\alpha\bpsi_\alpha + d\bxi_\alpha\psi^\alpha  \big) \big].
\eea
The forms for $su(2)$ generators  $T_\alpha{}^\beta$ and fermionic generators are
\bea\label{N4omegaOmega}
\big( \omega_T \big)_\beta{}^\alpha &=& -\im \left( e^{-\im\lambda} \right)_\gamma{}^\alpha d\left( e^{\im\lambda} \right)_\beta{}^\gamma + \left( e^{\im\lambda} \right)_\beta{}^\sigma\left( e^{-\im\lambda} \right)_\rho{}^\alpha\big( {\hat\omega}_T \big)_\sigma{}^\rho,  \\
\big(\omega_Q\big)^\alpha &=&\left( e^{-\im\lambda} \right)_\rho{}^\alpha \big(\hat\omega_Q\big)^\rho, \;\; \big(\omega_S\big)^\alpha =\left( e^{-\im\lambda} \right)_\rho{}^\alpha \big(\hat\omega_S\big)^\rho, \nn \\
\big(\bar\omega_Q\big)_\alpha &=&\left( e^{\im\lambda} \right)_\alpha{}^\rho \big(\hat{\bar\omega}_Q\big)_\rho, \;\;\big(\bar\omega_S\big)_\alpha =\left( e^{\im\lambda} \right)_\alpha{}^\rho \big(\hat{\bar\omega}_S\big)_\rho,\nn
\eea
where
\bea\label{2Ntildegmformsferm}
\big( \hat\omega_{T} \big)_\beta{}^\alpha &=& 2 \big(d\xi^\alpha \bpsi_\beta - d\bxi_\beta\psi^\alpha + \triangle t  \psi^\alpha \bpsi_\beta   \big)  - \delta_\beta^\alpha\big( d\xi^\gamma \bpsi_\gamma - d\bxi_\gamma\psi^\gamma + \triangle t  \psi^\gamma \bpsi_\gamma \big), \nn \\
\big( \hat\omega_{Q} \big)^\alpha &=& e^{-\frac{u}{2}} \big( d\xi^\alpha +\triangle t \psi^\alpha   \big), \;\; \big( \hat{\bar\omega}_{Q} \big)_\alpha = e^{-\frac{u}{2}} \big( d\bxi_\alpha +\triangle t \bpsi_\alpha  \big), \nn \\
\big( \hat\omega_{S} \big)^\alpha &=&e^{\frac{u}{2}}\big[  d\psi^\alpha +2\im d\bxi_\beta\psi^\beta \psi^\alpha -\im  \psi^\beta\bpsi_\beta d\xi^\alpha - \im\triangle t \psi^\alpha \, \psi^\beta\bpsi_\beta  +z \big( d\xi^\alpha + \triangle t \psi^\alpha  \big) \big],\\
\big( \hat{\bar\omega}_{S} \big)_\alpha &=&e^{\frac{u}{2}}\big[  d\bpsi_\alpha +2\im  d\xi^\beta\bpsi_\beta\, \bpsi_\alpha +\im  \psi^\beta\bpsi_\beta \, d\bxi_\alpha + \im\triangle t \bpsi_\alpha \, \psi^\beta\bpsi_\beta  +z \big( d\bxi_\alpha + \triangle t \bpsi_\alpha  \big)\big].\nn
\eea

We subject the forms to the usual conditions \p{susyidea}
\be\label{N4conds}
\omega_P = \triangle\tau, \;\; \big( \omega_Q \big)^\alpha =d\theta^\alpha, \;\; \big( \bar\omega_Q \big)_\alpha =d\bar\theta_\alpha, \;\; \omega_D =0,
\ee
where $\triangle\tau$, $d\theta^\alpha$, $d\bar\theta_\alpha$ are invariant with respect to $\cN{=}4$ supersymmetry transformations \footnote{Just as before, these expressions follow from the ``boundary''  Cartan forms defined  through the element  $g_0=e^{\im \tau P}\,e^{\theta^\alpha Q_\alpha + {\bar\theta}_\alpha\bQ^\alpha}$}
\be\label{N4deltatautheta}
\triangle\tau = d\tau + \im d\theta^\alpha \, \bar\theta_\alpha + \im d\bar\theta_\alpha \, \theta^\alpha, \;\; \delta \tau = \im \big( \epsilon^\alpha \bar\theta_\alpha + \bar\epsilon_\alpha\, \theta^\alpha  \big), \;\; \delta\theta^\alpha = \epsilon^\alpha, \;\; \delta\bar\theta_\alpha = \bar\epsilon_\alpha.
\ee

Correspondingly,  the $\cN{=}4$ covariant derivatives read
\bea\label{N4ders}
&&D_\alpha = \frac{\partial}{\partial \theta^\alpha}-\im \bar\theta_\alpha \partial_\tau, \;\; \bD^\alpha = \frac{\partial}{\partial \bar\theta_\alpha}-\im \theta^\alpha \partial_\tau, \nn \\ &&\big\{ D_\alpha, D_\beta \big\}=\big\{ \bD{}^\alpha, \bD{}^\beta \big\}=0, \;\; \big\{ D_\alpha, \bD{}^\beta \big\}=-2\im \delta_\alpha^\beta \partial_\tau.
\eea
From now on we will treat all fields as the superfields depending on the coordinates of ``boundary'' superspace $\left\{\tau, \theta^\alpha, {\bar\theta}_\alpha\right\}$.

The analysis of Maurer-Cartan equation, which we leave for Appendix C, shows that all the Cartan forms, aside of constrained $\omega_P$, $\big( \omega_Q\big)^\alpha$, $\big( \bar\omega_Q\big)_\alpha$, $\omega_D$, can be written in terms of three quantities $\cS_\alpha{}^\beta$, $\cS_\alpha{}^\alpha=0$:
\bea\label{N4formsfin}
\big(\omega_T\big)_\beta{}^\alpha &=& \cS_\beta{}^\alpha \triangle\tau,\;\; \big( \omega_S  \big)^\alpha = \frac{1}{3}\triangle \tau \bD^\gamma\cS_\gamma{}^\alpha - \im \cS_\beta{}^\alpha d\theta^\beta, \;\; \big( \bar\omega_S  \big)_\alpha = -\frac{1}{3}\triangle \tau D_\gamma\cS_\alpha{}^\gamma + \im \cS_\alpha{}^\beta d\bar\theta_\beta, \nn \\
\omega_K  &=& \triangle\tau \left( \frac{1}{12}\big[ D_\mu,\bD_\nu  \big]\cS^{\mu\nu} - \frac{1}{2}\cS_{\mu\nu}\cS^{\mu\nu}\right) - \frac{\im}{3}d\theta^\alpha \, D_\gamma\cS_\alpha{}^\gamma + \frac{\im}{3} d\bar\theta_\alpha \bD^\gamma\cS_\gamma{}^\alpha.
\eea
Unlike previously considered systems, $\cS_\alpha{}^\beta$ satisfy differential constraints
\be\label{N4vect}
D^{(\gamma}\cS^{\alpha\beta)} =0,\;\; \bD^{(\gamma}\cS^{\alpha\beta)} =0,
\ee
which imply that $\cS^{\alpha\beta}$ form an $\cN{=}4$, $d=1$ vector multiplet.
In full agreement with the previous cases we call $\cS^{\alpha\beta}$ by
$\cN{=}4$ super-Schwarzian
\be\label{N4S}
\cS_{\cN{=}4}^{\alpha\beta} = \cS^{\alpha\beta}.
\ee
The last step is to express $\cN{=}4$ super-Schwarzian in terms of our basic superfields $\left\{t, \xi^\alpha, \bxi_\alpha\right\}$.
The relations between the fields $t$, $\xi^\alpha$, $\bxi_\alpha$, $\psi^\alpha$, $\bpsi_\alpha$, $u$, $z$ and constraints on them can be obtained by expanding relations \p{N4conds} in terms of $\triangle\tau$, $d\theta^\alpha$, $d\bar\theta_\alpha$. The $\omega_P$ conditions read
\bea\label{N4sfcondsP}
\omega_P = \triangle\tau \;\; &\Rightarrow& \;\; \dot t+\im \big( \dot\xi{}^\alpha\, \bxi_\alpha + \dot\bxi_\alpha\xi^\alpha  \big) = e^u, \nn \\
&& D_\alpha t +\im \big( D_\alpha \xi{}^\beta\, \bxi_\beta + D_\alpha \bxi_\beta\xi^\beta  \big)=0, \nn \\  &&\bD{}^\alpha t +\im \big( \bD^\alpha \xi{}^\beta\, \bxi_\beta + \bD^\alpha \bxi_\beta\xi^\beta  \big)=0.
\eea
The $\big(\omega_Q\big)^\alpha$ and $\big(\bar\omega_Q\big)_\alpha$ conditions read
\bea\label{N4sfcondsQ}
\big(\omega_Q\big)^\alpha = d\theta^\alpha\; &\Rightarrow& \; \psi^\alpha = -e^{-u}  \dot\xi{}^\alpha, \;\; D_\beta\xi^\alpha =\big( e^{\im \lambda}  \big)_\beta{}^\alpha e^{u/2}, \;\; \bD^\beta \xi^\alpha=0, \\
\big(\bar\omega_Q\big)_\alpha = d\bar\theta_\alpha\; &\Rightarrow& \; \bpsi_\alpha = -e^{-u}\dot\bxi_\alpha , \;\; \bD{}^\beta\bxi_\alpha =\big( e^{-\im \lambda}  \big)_\alpha{}^\beta e^{u/2}, \;\; D_\beta \bxi_\alpha=0. \nn
\eea
Finally, the $\omega_D=0$ conditions are
\bea\label{N4sfcondsD}
\omega_D =0 \; &\Rightarrow& \; D_\alpha u = -2\im e^{-u} D_\alpha \xi^\beta \dot\bxi_{\beta}, \;\;  \bD{}^\alpha  u=  -2\im e^{-u} \bD{}^\alpha \bxi_\beta \dot\xi^{\beta} , \;\; z = \frac{1}{2}e^{-u}\dot u.
\eea
It can be shown that the conditions \p{N4sfcondsP}, \p{N4sfcondsQ} are equivalent to
\be\label{N4irrcond}
D_\alpha t + \im D_\alpha\xi^\gamma \, \bxi_\gamma =0, \;\; \bD^\alpha t+ \im \bD^\alpha \bxi_\gamma\, \xi^\gamma =0, \;\; D_\alpha\bxi_\beta=0, \;\; \bD^\alpha\xi^\beta =0
\ee
and define the $\cN{=}4$ linear multiplet, with one ``physical'' boson, four fermions and three ``auxiliary'' bosons (though in the system under discussion all are dynamical). It can be shown that the same phenomenon as in $\cN{=}2$ case happens: the commutator of covariant derivatives, acting on $t$, reduces to $\tau$-derivative of $\xi^\alpha$, $\bxi_\beta$:
\be\label{N4irrcond5}
\big[ D_\alpha, \bD{}^\beta \big]t = 2\delta_\alpha^\beta \partial_\tau\big(\xi^\mu \bxi_\mu  \big).
\ee
and $\xi^\alpha$, $\bxi_\beta$ can not be expressed entirely in terms of $t$. This again does not put the system on-shell:
\be\label{N4irrcond6}
D_\alpha \bD{}^\beta t = \delta_\alpha^\beta \big( -\im \dot t + \partial_\tau \big(\xi^\mu \bxi_\mu  \big) \big), \;\; D_\alpha D_\beta \bD{}^\beta t = -\frac{1}{2}D_\alpha D_\beta \bD{}^\beta t =0, \;\; \partial_\tau D_\alpha \big(- \im t + \xi^\mu \bxi_\mu \big)=0.
\ee
Now, one can obtain the $\cN{=}4$ Schwarzian as $\triangle\tau$ projection of the form $\big(\omega_T \big)_\alpha{}^\beta$. The mentioned projection reads
\bea\label{N4sfactomegaT}
\big(\omega_T\big)_\beta{}^\alpha =\triangle\tau\left[ -\im \big(   e^{-\im \lambda}\big)_\gamma{}^\alpha \partial_\tau \big(   e^{\im \lambda}\big)_\beta{}^\gamma - 2 e^{-u} \big(   e^{-\im \lambda}\big)_\mu{}^\alpha \big(   e^{\im \lambda}\big)_\beta{}^\nu \, \dot\xi{}^\mu\, \dot\bxi_\nu + \delta_\beta^\alpha \dot\xi{}^\mu\, \dot\bxi_\mu e^{-u}\right].
\eea
In comparison, calculating traceless part of $\big[ D_\beta,\bD{}^\alpha  \big]u$ using \p{N4sfcondsQ}, \p{N4sfcondsD}, one finds:
\bea\label{N4sfactDDu}
\left[ D_\beta,\bD{}^\alpha  \right]u - \frac{1}{2}\delta_\beta^\alpha \left[ D_\gamma,\bD{}^\gamma \right]u &=& -4\im \big(   e^{-\im \lambda}\big)_\gamma{}^\alpha \partial_\tau \big(   e^{\im \lambda}\big)_\beta{}^\gamma - \nn \\&& - 8 e^{-u} \big(   e^{-\im \lambda}\big)_\mu{}^\alpha \big(   e^{\im \lambda}\big)_\beta{}^\nu \, \dot\xi{}^\mu\, \dot\bxi_\nu + 4\delta_\beta^\alpha\, e^{-u} \, \dot\xi{}^\mu\, \dot\bxi_\mu .
\eea
Therefore, the $\cN{=}4$ Schwarzian reads
\bea\label{N4schw}
\big(\cS_{\cN{=}4} \big)_\beta{}^\alpha &=& \frac{1}{4} \left( \big[ D_\beta,\bD{}^\alpha  \big] - \frac{1}{2}\delta_\beta^\alpha \big[ D_\gamma,\bD{}^\gamma \big]   \right)u =\nn \\&=& \frac{1}{4} \big( \big[ D_\beta,\bD{}^\alpha  \big] - \frac{1}{2}\delta_\beta^\alpha \big[ D_\gamma,\bD{}^\gamma \big]   \big)\log \big( D_\mu\xi^\nu \, \bD^\mu \bxi_\nu  \big),
\eea
as expected.

As $\cS^{\alpha\beta}$ satisfies the irreducibility conditions of the vector multiplet, $\big[ D_\mu,\bD_\nu  \big]S^{\mu\nu}$ transforms w.r.t. supersymmetry as an auxiliary field (gets shifted by a total derivative). Therefore,
\bea\label{N4sfact}
S_{N4schw} &=&-\frac{1}{12} \int d\tau \big[ D_\mu,\bD_\nu  \big]\cS^{\mu\nu} =\frac{1}{6} \int d\tau d\theta^\alpha d\bar\theta_\beta\, \cS_\alpha{}^\beta
\eea
is the $\cN{=}4$ Schwarzian action. In terms of Cartan forms, it can be presented as
\bea\label{N4sfactform}
S_{N4schw}&=& \frac{1}{6} \int \big(\omega_Q\big)^\alpha\, \wedge\, \big({\bar\omega}_Q \big)_\beta\, \wedge\,\big(\omega_T\big)_\alpha{}^\beta = \frac{\im}{6} \int \omega_P \, \wedge \, \big(\omega_S\big)^\alpha\, \wedge\, \big({\bar\omega}_Q \big)_\alpha=\nn \\& =& - \frac{\im}{6} \int \omega_P \, \wedge \, \big(\omega_Q\big)^\alpha\, \wedge\, \big({\bar\omega}_S \big)_\alpha.
\eea
Evaluating the integral in \p{N4sfact}, one can obtain the component action
\bea\label{N4compact}
S_{N4schw}= - \frac{1}{2}\int d\tau \left[ \frac{\partial_\tau^2 \big( \dot t + \im \dot\xi^\alpha \bxi_\alpha +  \im \dot\bxi_\alpha \xi^\alpha \big)}{\dot t + \im \dot\xi^\alpha \bxi_\alpha +  \im \dot\bxi_\alpha \xi^\alpha} - \frac{3}{2}\left( \frac{\partial_\tau \big( \dot t + \im \dot\xi^\alpha \bxi_\alpha +  \im \dot\bxi_\alpha \xi^\alpha \big)}{\dot t + \im \dot\xi^\alpha \bxi_\alpha +  \im \dot\bxi_\alpha \xi^\alpha}  \right)^2 + \right. \nn \\
\left. +2 \im \frac{\ddot\xi{}^\alpha \dot\bxi_\alpha + \ddot\bxi_\alpha \dot\xi{}^\alpha}{\dot t + \im \dot\xi^\beta \bxi_\beta +  \im \dot\bxi_\beta \xi^\beta}  + \big( e^{-\im \lambda}   \big)_\rho{}^\beta \, \big( e^{-\im \lambda}   \big)_\sigma{}^\alpha \partial_\tau \big( e^{\im \lambda}   \big)_\alpha{}^\rho \, \partial_\tau \big( e^{\im \lambda}   \big)_\beta{}^\sigma  -  \right. \nn \\
\left.  -4\im \frac{\big( e^{-\im \lambda}   \big)_\rho{}^\beta \partial_\tau \big( e^{\im \lambda}   \big)_\beta{}^\sigma \dot\xi{}^\rho \dot\bxi_\sigma }{\dot t + \im \dot\xi^\alpha \bxi_\alpha +  \im \dot\bxi_\alpha \xi^\alpha} \right].
\eea
Here, as usual, $t$, $\xi{}^\alpha$, $\bxi_\alpha$, $\lambda_\alpha{}^\beta$ are the first components of respective superfields.

\setcounter{equation}{0}
\section{Conclusion}
In this work we re-consider the application of the method of nonlinear realizations to the $\cN=0,1,2,3,4$ (super)conformal  groups. As compared to the previous attempts to utilize the nonlinear realizations for construction of the super-Schwarzians \cite{AG1,AG2,AG3,AG4,AG5}, our consideration is based on the minimal set of constraints imposed on the Cartan forms. These constraints include
\begin{itemize}
	\item The constraints on the forms of $\cN$-extended super Poincare generators $\omega_P=\triangle \tau, \omega^i_Q = d\theta^i$.
	Here, the forms $\triangle \tau, d\theta^i$ depend on the coordinates of  ``boundary'' superspace $\left\{\tau, \theta^i\right\}$  and they are invariant with respect to rigid $\cN$-extended supersymmetry transformations;
	\item The final constraint reads $\omega_D=0$. It provides some variant of the Inverse Higgs Phenomenon constraints \cite{ih}.
\end{itemize}
We explicitly show that this minimal set of constraints is enough to express
all Cartan forms of the $\cN=0,1,2,3,4$ (super)conformal  groups in terms of corresponding (super-)Schwarzians and their derivatives.

In the cases of higher supersymmetries the calculations quickly become rather cumbersome. Having at hands the constraints written on the Cartan forms (not on their projections!) it proved useful to use the Maurer-Cartan equations which
help to express all Cartan forms in terms of the single object - $\cN$ super-Schwarzian. However, to find the expression of the  $\cN$ super-Schwarzian
in terms of the basic superfields one has to again use all set of constraints.

The idea to use the ``boundary'' superspace to impose the proper constrains on the Cartan forms was firstly formulated in \cite{AG1,AG2}. However, the full set of constraints used in the papers \cite{AG1,AG2,AG3,AG4,AG5} seems to be unessentially strong. At least our analysis shows that these constraints unavoidably restrict super-Schwarzians.

We are planning to apply the proposed approach to $\cN$-extended superconformal group including the variant of $OSp(4|2)$ superconformal symmetry. Another
interesting problem is to obtain non-relativistic and/or Carrollian versions of
the Schwarzian \cite{gomis}.

\acknowledgments
The work was supported by Russian Foundation for Basic Research, grant
No~20-52-12003.

\appendix
\setcounter{equation}{0}
\section{Maurer-Cartan equations for $SU(1,1|1)$}

Let us demonstrate the usefulness of the Maurer-Cartan equations for supergroup $SU(1,1|1)$ on the example of $\cN{=}2$ super Schwarzian.

We find it preferable to write down the Maurer-Cartan equation in the form :
\be\label{mainMCequation}
d_2 \Omega(d_1) - d_1 \Omega(d_2) = \left[ \Omega(d_1), \Omega(d_2)\right].
\ee
Here, differentials $d_1$, $d_2$ are assumed to commute, $d_1\, d_2 = d_2\, d_1$, and differentials of bosonic and fermionic functions are bosons and fermions, respectively. If $\Omega(d_i) = g^{-1}d_i g$, as it should be for a Cartan form, equation \p{mainMCequation} reduces to just an identity. However, if one substitutes $\Omega$ just as in \p{N2cfdef}, it would be possible to derive nontrivial relations the structure functions of the forms satisfy.

Substituting the expansion of the Cartan form in generators \p{N2cfdef} into \p{mainMCequation}, one obtains
\bea\label{N2MCP}
d_2 \omega_{1P} -d_1\omega_{2P} &=& -\big( \omega_{1P}\omega_{2D}- \omega_{1D}\omega_{2P} \big)+2\im \big(\omega_{1Q} \bar\omega_{2Q}  +\bar\omega_{1Q}   \omega_{2Q} \big), \\ \label{N2MCK}
d_2 \omega_{1K} -d_1\omega_{2K}  &=&  \big( \omega_{1K}\omega_{2D}- \omega_{1D}\omega_{2K} \big)+2\im\big( \omega_{1S} \bar\omega_{2S} + \bar\omega_{1S}  \omega_{2S} \big), \\ \label{N2MCD}
d_2 \omega_{1D} -d_1\omega_{2D}  &=& -2 \big( \omega_{1P}\omega_{2K}- \omega_{1K}\omega_{2P} \big)-\nn \\&&-2\im\big( \omega_{1Q} \bar\omega_{2S}  +\bar\omega_{1Q} \omega_{2S}+ \omega_{1S} \bar\omega_{2Q}  + \bar\omega_{1S} \omega_{2Q} \big),  \\ \label{N2MCJ}
d_2 \omega_{1J} -d_1\omega_{2J} &=& -2\big(  \omega_{1Q}  \bar\omega_{2S}  - \bar\omega_{1Q}  \omega_{2S}- \omega_{1S}  \bar\omega_{2Q}  + \bar\omega_{1S}  \omega_{2Q} \big),  \\\label{N2MCQ}
d_2 \omega_{1Q} -d_1\omega_{2Q} &=& \omega_{1P}   \omega_{2S}  -  \omega_{2P}  \omega_{1S}  + \frac{1}{2}\big( \omega_{1D}   \omega_{2Q} -  \omega_{2D}  \omega_{1Q}    \big) -\nn \\&& - \frac{\im}{2}\big( \omega_{1J}   \omega_{2Q}  - \omega_{2J}   \omega_{1Q}   \big),  \\ \label{N2MCQb}
d_2 \bar\omega_{1Q} -d_1 \bar\omega_{2Q} &=& \omega_{1P}  \bar\omega_{2S}  -  \omega_{2P}   \bar\omega_{1S} + \frac{1}{2}\big( \omega_{1D} \bar\omega_{2Q}  -  \omega_{2D}   \bar\omega_{1Q}   \big) +\nn \\&&+ \frac{\im}{2}\big( \omega_{1J}   \bar\omega_{2Q}  - \omega_{2J}   \bar\omega_{1Q}   \big),  \\ \label{N2MCS}
d_2\omega_{1S}-d_1\omega_{2S} &=& -\omega_{1K}   \omega_{2Q} +  \omega_{2K}   \omega_{1Q} - \frac{1}{2}\big( \omega_{1D}  \omega_{2S} -  \omega_{2D}  \omega_{1S}   \big) -\nn \\ &&- \frac{\im}{2}\big( \omega_{1J}   \omega_{2S} - \omega_{2J}   \omega_{1S} \big),  \\ \label{N2MCSb}
d_2 \bar\omega_{1S} -d_1 \bar\omega_{2S} &=& -\omega_{1K} \bar\omega_{2Q} +  \omega_{2K}  \bar\omega_{1Q}  - \frac{1}{2}\big( \omega_{1D}   \bar\omega_{2S}  -  \omega_{2D}   \bar\omega_{1S}    \big) +\nn \\&&+ \frac{\im}{2}\big( \omega_{1J}   \bar\omega_{2S}  - \omega_{2J}   \bar\omega_{1S}   \big) .
\eea
Here, to make notation shorter, we denote $\omega_{1P} = \omega_P (d_1)$ and so on. Explicit substitution $\omega_{1P} = e^{-u}\big(   d_1 t+ \im d_1\xi\bxi +\im d_1\bxi\xi \big)$ and others should reduce these equations to identities. Let us, however, impose the constraints \p{eqN2t}, \p{eqN2xi}, \p{eqN2bxi},  and $\omega_D=0$ directly on the forms. Then all the forms \p{N2cfdef} should written in terms of $\triangle\tau$, $d\theta$, $d\bar\theta$:
\bea\label{N2omegas}
\omega_P&=& \triangle\tau, \;\; \omega_Q = d\theta, \;\; \bar\omega_Q = d\bar\theta, \;\; \omega_D =0, \;\; \omega_J = \im \triangle\tau \cS + d\theta\, \Phi - d\bar\theta {\overline \Phi},  \\
\omega_S &=& \triangle\tau \, \Psi + d\theta A + d\bar\theta B, \;\; \bar\omega_S = \triangle\tau \, {\overline\Psi} + d\theta {\overline B} + d\bar\theta {\overline A}, \;\;  \omega_K = \triangle\tau C + d\theta \Sigma -d\bar\theta {\overline \Sigma},\nn
\eea
where $\cS$, $\Phi$, ${\overline \Phi}$, $A$, ${\overline A}$, $B$, ${\overline B}$, $C$, $\Sigma$, ${\overline \Sigma}$ are so far unconstrained superfunctions.

With $\omega_P = \triangle\tau$ and $\omega_Q = d\theta$, ${\bar\omega}_Q = d\bar\theta$, $d\omega_P$ equation \p{N2MCP} is satisfied identically, as
\bea\label{N2MCP2}
&&d_2\triangle_1 \tau - d_1 \triangle_2 \tau = d_2 \big(  d_1 \tau +\im \big(d_1 \theta \bar\theta + d_1\bar\theta \theta  \big)\big) - d_1 \big(  d_2 \tau +\im \big(d_2 \theta \bar\theta + d_2\bar\theta \theta  \big)\big) = \nn \\
&&=\big(d_2d_1-d_1d_2 \big)\tau - \im \big(d_2d_1-d_1d_2 \big)\theta \, \bar\theta - \im \big(d_2d_1-d_1d_2 \big)\bar\theta \, \theta +\nn \\&&+2\im d_1\theta \, d_2\bar\theta +  2\im d_1\bar\theta d_2\theta =
=2\im d_1\theta \, d_2\bar\theta +  2\im d_1\bar\theta d_2\theta,
\eea
as the differentials $d_1$, $d_2$ commute. Note that as $\omega_D =0$, equation \p{N2MCP} is just the Maurer-Cartan equation satisfied by the Cartan forms of $\cN{=}2$, $d=1$ Poincare supergroup. Therefore, the choice of conditions $\omega_P = \triangle\tau$ and $\omega_Q = d\theta$, ${\bar\omega}_Q = d\bar\theta$, where $\triangle\tau$, $d\theta$, $d\bar\theta$ is are standard invariant forms on $\cN{=2}$, $d=1$ superspace, is rather natural from supergeometry point of view.

Substituting this relation into equation \p{N2MCQ}, we find that
\bea\label{N2MCQ2}
0 = d_2d_1\theta - d_1d_2\theta &=& \big( \triangle_1 \tau \, d_2\theta - \triangle_2\tau d_1\theta  \big)\left(   A + \frac{1}{2}\cS \right) + \big( \triangle_1 \tau \, d_2\bar\theta - \triangle_2\tau d_1\bar\theta  \big)B+\nn \\&&+\im d_1\theta\, d_2\theta \Phi + \frac{\im}{2} \big( d_1\bar\theta\, d_2\theta - d_2\bar\theta \, d_1\theta  \big){\overline \Phi}.
\eea
While $\Psi$ is yet undetermined, just one equation \p{N2MCQ} is strong enough to show that the form $\omega_S$ can not have a $d\bar\theta$ - projection, and $d\theta$ and $d\bar\theta$ projections of $\omega_J$ are absent. Also it relates $d\theta$ projection of $\omega_S$ and $\triangle\tau$ projection of $\omega_J$: $A = -1/2\cS$. The analysis of $d\bar\omega_Q$ equation \p{N2MCQb} leads to analogous results ${\overline B}=0$, ${\overline A} = 1/2\cS$.

Most convenient next step would be to study $d\omega_J$ equation \p{N2MCJ}. As we already reduced $\omega_J$ to $\omega_J = \im \triangle\tau \cS$, taking into account that
\be\label{N2MCJ1}
d_2 \omega_{1J} = d_2\triangle_1\tau\, \cS + \triangle_1\tau \big( \triangle_2 \tau \, \dot{\cS} + d_2\theta D\cS +d_2\bar\theta \, \bD\cS \big),
\ee
we find
\be\label{N2MCJ2}
d_2 \omega_{1J} -d_1\omega_{2J} =-2\big( d_1\theta \, d_2\bar\theta + d_1\bar\theta d_2\theta  \big)\cS + \im\big( \triangle_1 \tau d_2\theta - \triangle_2 \tau d_1\theta \big)D\cS + \im\big( \triangle_1 \tau d_2\bar\theta - \triangle_2 \tau d_1\bar\theta \big)\bD\cS.
\ee
Comparing this with the right hand side of \p{N2MCJ}, where $\omega_S = \triangle\tau \Psi -1/2 d\theta \cS$ and $\bar\omega_S = \triangle\tau {\overline \Psi} +1/2 d\bar\theta \cS$, we find that $d\theta\times d\bar\theta$ terms cancel from \p{N2MCJ}, while the rest imply
\be\label{N2MCJ3}
\Psi = -\frac{\im}{2}\bD\cS, \;\; {\overline\Psi} = \frac{\im}{2}D\cS,
\ee
and the forms $\omega_S$, $\bar\omega_S$, $\omega_J$ can be written in terms of just one quantity $\cS$:
\be\label{N2omegaSJres}
\omega_S =  -\frac{\im}{2}\triangle\tau \bD\cS - \frac{1}{2}d\theta\,\cS, \;\; {\bar\omega}_S =  \frac{\im}{2}\triangle\tau \cS + \frac{1}{2}d\bar\theta\,\cS, \;\; \omega_J = \im \triangle\tau \cS.
\ee
Next step further would be to check $d\omega_D$ equation \p{N2MCD}. As the left hand side of \p{N2MCD} is zero due condition $\omega_D=0$, we do not need to take the differential of anything. Simply putting results for the forms $\omega_S$, $\bar\omega_S$ and the ansatz for $\omega_K$ into \p{N2MCD}, we find that $d\theta\times d\theta$ terms cancel and others imply
\be\label{N2Sigmasol}
\Sigma = -\frac{1}{2}D\cS, \;\; {\overline \Sigma} = -\frac{1}{2}\bD\cS.
\ee
To determine $C$, we should consider $d\omega_S$ or $d\bar\omega_S$ equations \p{N2MCS}, \p{N2MCSb}. The left hand side of  \p{N2MCS} can be calculated from \p{N2omegaSJres} as
\bea\label{N2MCS1}
d_2 \omega_{1S} - d_1\omega_{2S} &=& -d_1\theta\, d_2\theta D\cS + \frac{1}{2} \big(d_1\theta d_2\bar\theta + d_1\bar\theta d_2\theta  \big)\bD\cS + \nn \\&&+\big( \triangle_1 \tau d_2\bar\theta - \triangle_2 \tau d_1\bar\theta  \big) \left( -\frac{\im}{2}D\bD\cS +\frac{1}{2}\dot\cS  \right).
\eea
Substituting the $\omega_S$, $\bar\omega_S$ and $\omega_J$ \p{N2omegaSJres} to the right hand side, as well as the ansatz for $\omega_K$ \p{N2omegas}, one obtains
\bea\label{N2MCS2}
&&-\omega_{1K}   \omega_{2Q} +  \omega_{2K}   \omega_{1Q} - \frac{\im}{2}\big( \omega_{1J}   \omega_{2S} - \omega_{2J}   \omega_{1S} \big) = \\&&= -d_1\theta\, d_2\theta D\cS + \frac{1}{2} \big(d_1\theta d_2\bar\theta + d_1\bar\theta d_2\theta  \big)\bD\cS+ \big( \triangle_1 \tau d_2\bar\theta - \triangle_2 \tau d_1\bar\theta  \big) \left( -C -\frac{1}{4}\cS^2 \right).\nn
\eea
Thus all $d\theta \times d\theta$, $d\theta\times d\bar\theta$ terms cancel out and one finds
\be\label{N2omegaKsol}
C = \frac{\im}{4}\big[ D,\bD \big]\cS - \frac{1}{4}\cS^2, \;\; \omega_K = \frac{1}{4}\triangle\tau \big(\im\big[ D,\bD \big]\cS -\cS^2 \big) -\frac{1}{2} d\theta D\cS +\frac{1}{2}d\bar\theta \bD\cS.
\ee
As every projection of all the forms is already found in terms of $\cS$ and its derivatives, one can only check by direct calculation that $d\omega_K$ equation \p{N2MCK} is satisfied. It is indeed so, with no constraints imposed on $\cS$.

The results \p{N2omegaSJres}, \p{N2omegaKsol} are in full agreement with ones obtained by straightforward calculation of \p{finformsN2}. Though for $\cN{=}2$ Schwarzians this analysis was somewhat tedious and not particularly easier than direct calculation of multiplet defining conditions, it is still important. At first, it shows that the structure of Cartan forms in $\cN{=1}$ and $\cN{=}2$ cases is not a coincidence and reflects fundamental properties of supersymmetric Schwarzians. Secondly, in the $\cN{=}3$ and $\cN{=}4$ cases, the irreducibility conditions of multiplets become more and more important, while remaining highly nonlinear, and calculation of their consequences becomes increasingly difficult. Therefore, analysis of Maurer-Cartan equations becomes more convenient way to identify proper Schwarzians even from technical point of view.

\setcounter{equation}{0}
\section{Maurer-Cartan equations for $OSp(3|2)$}

The main Maurer-Cartan equation
\be
d_2\Omega_1 - d_1 \Omega_2 = \big[ \Omega_1, \Omega_2  \big] \nn
\ee
for $osp(3|2)$ with general structure of the Cartan form given by \p{N3cfdef} can be reduced to a set of relations
\bea\label{N3MC}
\im \left( d_2 \omega_{1P}- d_1 \omega_{2P} \right)  & = &  - 2 \big(\omega_{1Q}\big)_i \big(\omega_{2Q}\big)_i + \im \big( \omega_{1D}\omega_{2P}  - \omega_{2D}\omega_{1P} \big), \nn\\
d_2 \big(\omega_{1Q}\big)_i- d_1 \big(\omega_{2Q}\big)_i & = &  \omega_{1P}\big(\omega_{2S}\big)_i-\omega_{2P}\big(\omega_{1S}\big)_i-\frac{1}{2}\omega_{2D} \big(\omega_{1Q}\big)_i+\frac{1}{2}\omega_{1D} \big(\omega_{2Q}\big)_i+\nn \\&&+\epsilon_{imn}\left( \big(\omega_{1Q}\big)_m \big(\omega_{2J}\big)_n -\big(\omega_{2Q}\big)_m \big(\omega_{1J}\big)_n\right),  \\
\im \left( d_2 \omega_{1D}- d_1 \omega_{2D} \right)  & = &  - 2 \im \left( \omega_{1P}\omega_{2K} - \omega_{2P}\omega_{1K}\right) +2\left( \big(\omega_{1Q}\big)_i\big(\omega_{2S}\big)_i-\big(\omega_{2Q}\big)_i\big(\omega_{1S}\big)_i\right) , \nn\\
\im \left( d_2 \big(\omega_{1J}\big)_i- d_1 \big(\omega_{2J}\big)_i \right) & = & \epsilon_{imn}\left( \big(\omega_{1Q}\big)_m \big(\omega_{2S}\big)_n -\big(\omega_{2Q}\big)_m \big(\omega_{1S}\big)_n +\im \big(\omega_{1J}\big)_m\big(\omega_{2J}\big)_n\right), \nn \\
\left( d_2 \big(\omega_{1S}\big)_i- d_1 \big(\omega_{2S}\big)_i \right) & = &   \omega_{2K}\big(\omega_{1Q}\big)_i-\omega_{1K}\big(\omega_{2Q}\big)_i+\frac{1}{2}\omega_{2D} \big(\omega_{1Q}\big)_i-\frac{1}{2}\omega_{1D} \big(\omega_{2Q}\big)_i+\nn \\&&+\epsilon_{imn}\left( \big(\omega_{1J}\big)_m \big(\omega_{2S}\big)_n-\big(\omega_{2J}\big)_m \big(\omega_{1S}\big)_n\right), \nn \\
\im \left( d_2 \omega_{1K}- d_1 \omega_{2K} \right)  & = &  - 2 \big(\omega_{1S}\big)_i \big(\omega_{2S}\big)_i-\im \big( \omega_{1D}\omega_{2K} - \omega_{2D}\omega_{1K} \big).\nn
\eea
The primary conditions are $\omega_P =\triangle \tau$, $\big(\omega_Q \big)_i = d\theta_i$ and $\omega_D =0$. The remaining forms can be expanded in terms of $\triangle\tau$ and $d\theta_i$ as
\be\label{anzJ}
\big(\omega_J\big)_i = \triangle\tau B_i + d\theta_j\, \cS_{ij}, \;\; \big( \omega_S \big)_i = \triangle \tau \,\Psi_i + A_{ij} d\theta_j, \;\; \omega_K = \triangle\tau\, C + d\theta_i \Sigma_i.
\ee

Analyzing equations \p{N3MC}, one can obtain that $d\omega_P$ equation is satisfied automatically. The part of $d\omega_Q$ equation, proportional to $\triangle \tau \wedge d\theta$, implies that $A_{ij} = \epsilon_{ijk}B_k$, and $d\theta \wedge d\theta$ part is satisfied if $\epsilon_{ijk}\cS_{jm} + \epsilon_{ijm}\cS_{jk} =0$, which happens if and only if $\cS_{ij} = \delta_{ij}\cS$.

Analyzing $d\omega_J$ equation one obtains from $d\theta \wedge d\theta$ part that $B_j = \im D_{j}\cS$, while the $\triangle \tau \wedge d\theta$ part implies that
\be\label{bigpsi}
\Psi_i = \cS \, D_i \cS - \frac{1}{2}\epsilon_{ijk}D_j D_k \cS.
\ee
The $d\theta \wedge d\theta$ part of $d\omega_D$ equation is then satisfied automatically, and the rest implies just $\Sigma_i = \im \Psi_i$. The $d\omega_S$ equation is more complicated, producing two relations:
\bea\label{domegaS}
&&\big( \triangle_1 \tau\, d_2\theta_j -  \triangle_2 \tau\, d_1\theta_j   \big) \big( D_j \Psi_i - \im \epsilon_{ijk}D_k \dot\cS + C \delta_{ij} + \delta_{ij} B_k B_k - B_i B_j + \epsilon_{ijk}\cS\,\Psi_k   \big)=0, \nn \\
&&2\im \big( d_1 \theta_j d_2\theta_j \big) \Psi_i + \im \epsilon_{ijk} (d_1\theta_j d_2\theta_m + d_1\theta_m d_2\theta_j \big)D_m D_k \cS =\\&&= \im (d_1\theta_i d_2\theta_m + d_1\theta_m d_2\theta_i\big)\Psi_m - (d_1\theta_i d_2\theta_j + d_1\theta_j d_2\theta_i \big)\cS B_j +2 \big( d_1 \theta_j d_2\theta_j \big) \cS B_i. \nn
\eea
Substituting $B_i$, $\Psi_i$ into these equations, one finds that of the first one only $\delta_{ij}$ component survives while the second one is satisfied automatically. To show this, one should use the identity
\bea\label{epsid}
&&\epsilon_{ijk}X_m = \epsilon_{mjk}X_i + \epsilon_{imk}X_j + \epsilon_{ijm}X_k \; \Rightarrow \nn \\
&&\; D_m \big(\epsilon_{ipq} D_p D_q \cS  \big) = -2\im \epsilon_{imk}D_k\dot\cS + \frac{1}{3}\delta_{im} \big(\epsilon_{pqr} D_p D_q D_r \cS  \big).
\eea
After that, one obtains
\be\label{N3C}
C = -\im \cS \dot\cS + \frac{1}{6}\big(\epsilon_{pqr} D_p D_q D_r \cS  \big) - D_k \cS D_k\cS.
\ee
The $d\omega_K$ equation reduces to two relations
\be\label{domegaK}
\im D_i C + \dot\Psi_i + 2\im \epsilon_{ijk}\Psi_j D_k \cS =0, \;\; -2 \im \delta_{ij} C - D_j \Psi_i - D_i\Psi_j -2 \delta_{ij}D_k\, \cS D_k \cS+ 2D_i \cS\, D_j\cS=0.
\ee
They are satisfied identically, leaving no constraints on $\cS$. To prove this, one should use the relation $D_i \big(\epsilon_{pqr} D_p D_q D_r \cS  \big) = -3\im \epsilon_{ipq} D_p D_q  \dot\cS$, which follows from \p{epsid}.
Thus one obtains the complete solution of $osp(3|2)$ Maurer-Cartan equations \p{N3formssol}.

\setcounter{equation}{0}
\section{Maurer-Cartan equations for $SU(1,1|2)$}
With $\Omega$ given by \p{N4cfdef}, the Maurer-Cartan equation
\be
d_2\Omega_1 - d_1 \Omega_2 = \big[ \Omega_1, \Omega_2  \big] \nn
\ee
 splits into bosonic equations
\bea\label{2NMCbos}
\im \big( d_2 \omega_{1P} -d_1\omega_{2P} \big) &=& -\im \big( \omega_{1P}\omega_{2D}- \omega_{1D}\omega_{2P} \big)-2\big( \omega_{1Q}\big)^\alpha \big( \bar\omega_{2Q}  \big)_\alpha-2\big( \bar\omega_{1Q}  \big)_\alpha \big( \omega_{2Q}\big)^\alpha, \nn \\
\im \big( d_2 \omega_{1K} -d_1\omega_{2K} \big) &=& \im \big( \omega_{1K}\omega_{2D}- \omega_{1D}\omega_{2K} \big)-2\big( \omega_{1S}\big)^\alpha \big( \bar\omega_{2S}  \big)_\alpha-2\big( \bar\omega_{1S}  \big)_\alpha \big( \omega_{2S}\big)^\alpha,\nn \\
\im \big( d_2 \omega_{1D} -d_1\omega_{2D} \big) &=& -2\im \big( \omega_{1P}\omega_{2K}- \omega_{1K}\omega_{2P} \big)+2\big( \omega_{1Q}\big)^\alpha \big( \bar\omega_{2S}  \big)_\alpha+\nn \\&&+2\big( \bar\omega_{1Q}  \big)_\alpha \big( \omega_{2S}\big)^\alpha+ 2\big( \omega_{1S}\big)^\alpha \big( \bar\omega_{2Q}  \big)_\alpha+2\big( \bar\omega_{1S}  \big)_\alpha \big( \omega_{2Q}\big)^\alpha,  \\
d_2 \big(\omega_{1T}\big)_\beta{}^\alpha -d_1 \big(\omega_{2T}\big)_\beta{}^\alpha &=&  2\big(  \big( \omega_{1Q}\big)^\alpha \big( \bar\omega_{2S}  \big)_\beta-\big( \bar\omega_{1Q}  \big)_\beta \big( \omega_{2S}\big)^\alpha -\big( \omega_{1S}\big)^\alpha \big( \bar\omega_{2Q}  \big)_\beta+ \nn \\&&+\big( \bar\omega_{1S}  \big)_\beta \big( \omega_{2Q}\big)^\alpha \big)  -  \delta_\beta^\alpha \big(  \big( \omega_{1Q}\big)^\gamma \big( \bar\omega_{2S}  \big)_\gamma -\nn \\&&-\big( \bar\omega_{1Q}  \big)_\gamma \big( \omega_{2S}\big)^\gamma-\big( \omega_{1S}\big)^\gamma \big( \bar\omega_{2Q}  \big)_\gamma+\big( \bar\omega_{1S}  \big)_\gamma \big( \omega_{2Q}\big)^\gamma \big) + \nn \\
&&+  \im \big(\omega_{1T}\big)_\mu{}^\alpha \big(\omega_{2T}\big)_\beta{}^\mu -  \im \big(\omega_{1T}\big)_\beta{}^\mu \big(\omega_{2T}\big)_\mu{}^\alpha.\nn
\eea
and fermionic equations:
\bea\label{2NMCferm}
d_2 \big(\omega_{1Q}\big)^\alpha -d_1 \big(\omega_{2Q}\big)^\alpha &=& \omega_{1P}  \big( \omega_{2S} \big)^\alpha -  \omega_{2P}  \big( \omega_{1S} \big)^\alpha + \frac{1}{2}\big( \omega_{1D}  \big( \omega_{2Q} \big)^\alpha -  \omega_{2D}  \big( \omega_{1Q} \big)^\alpha   \big) + \nn \\
&&  + \im \big( \big(\omega_{1T}\big)_\beta{}^\alpha \big(\omega_{2Q} \big)^\beta - \big(\omega_{2T}\big)_\beta{}^\alpha \big( \omega_{1Q} \big)^\beta \big), \nn \\
d_2 \big(\bar\omega_{1Q}\big)_\alpha -d_1 \big(\bar\omega_{2Q}\big)_\alpha &=& \omega_{1P}  \big( \bar\omega_{2S} \big)_\alpha -  \omega_{2P}  \big( \bar\omega_{1S} \big)_\alpha + \frac{1}{2}\big( \omega_{1D}  \big( \bar\omega_{2Q} \big)_\alpha -  \omega_{2D}  \big( \bar\omega_{1Q} \big)_\alpha   \big) - \nn \\
&&  - \im \big( \big(\omega_{1T}\big)_\alpha{}^\beta \big(\bar\omega_{2Q} \big)_\beta - \big(\omega_{2T}\big)_\alpha{}^\beta \big(\bar\omega_{1Q} \big)_\beta \big), \\
d_2 \big(\omega_{1S}\big)^\alpha -d_1 \big(\omega_{2S}\big)^\alpha &=& -\omega_{1K}  \big( \omega_{2Q} \big)^\alpha +  \omega_{2K}  \big( \omega_{1Q} \big)^\alpha - \frac{1}{2}\big( \omega_{1D}  \big( \omega_{2S} \big)^\alpha -  \omega_{2D}  \big( \omega_{1S} \big)^\alpha   \big)+ \nn \\
&&  + \im \big( \big(\omega_{1T}\big)_\beta{}^\alpha \big(\omega_{2S} \big)^\beta - \big(\omega_{2T}\big)_\beta{}^\alpha \big( \omega_{1S} \big)^\beta \big), \nn \\
d_2 \big(\bar\omega_{1S}\big)_\alpha -d_1 \big(\bar\omega_{2S}\big)_\alpha &=& -\omega_{1K}  \big( \bar\omega_{2Q} \big)_\alpha+  \omega_{2K}  \big( \bar\omega_{1Q} \big)_\alpha - \frac{1}{2}\big( \omega_{1D}  \big( \bar\omega_{2S} \big)_\alpha -  \omega_{2D}  \big( \bar\omega_{1S} \big)_\alpha   \big) - \nn \\
&& - \im \big( \big(\omega_{1T}\big)_\alpha{}^\beta \big(\bar\omega_{2S} \big)_\beta - \big(\omega_{2T}\big)_\alpha{}^\beta \big(\bar\omega_{1S} \big)_\beta \big). \nn
\eea
With conditions \p{N4conds} applied and all the forms written as combinations of $\triangle\tau$, $d\theta^\alpha$, $d\bar\theta_\alpha$ with superfield coefficients
\bea\label{N4formsans}
\big(\omega_S\big)^\alpha = \triangle\tau\, \lambda^\alpha +d\theta^\beta A_\beta{}^\alpha + d\bar\theta_\beta B^{\alpha\beta}, \;\; \big(\bar\omega_S\big)_\alpha = \triangle\tau\, \bar\lambda_\alpha + d\bar\theta_\beta\, {\overline A}_{\alpha}{}^\beta + d\theta^\beta {\overline B}_{\alpha\beta}, \nn \\
\big( \omega_T  \big)_\alpha{}^\beta = \cS_\alpha{}^\beta \triangle\tau + d\theta^\gamma\, \Sigma_{\gamma\alpha}{}^\beta - d\bar\theta_\gamma\, {\overline \Sigma}^\gamma{}_\alpha{}^\beta, \;\;
\omega_K = \triangle\tau \,C + d\theta^\alpha\chi_\alpha - d\bar\theta_\alpha\bar\chi^\alpha,
\eea
it can be straightforwardly checked that $d\omega_P$ equation \p{2NMCbos} is satisfied identically. Substituting \p{N4formsans} into the $d\omega_Q$ equation, one finds
\bea\label{N4dQeq}
0=\big(\triangle_1\tau d_2\bar\theta_\beta - \triangle_2\tau d_1\bar\theta_\beta   \big)B^{\alpha\beta} + \big(\triangle_1\tau d_2\theta^\beta - \triangle_2\tau d_1\theta^\beta   \big)\big( A_\beta{}^\alpha+\im S_\beta{}^\alpha  \big) + \nn \\
+\im d_1\theta^\gamma d_2\theta^\beta \big( \Sigma_{\gamma\beta}{}^\alpha - \Sigma_{\beta\gamma}{}^\alpha  \big) - \im \big(d_1\bar\theta_\gamma d_2\theta^\beta -d_2\bar\theta_\gamma d_1\theta^\beta    \big){\overline \Sigma}^\gamma{}_\beta{}^\alpha =0.
\eea
Therefore, one should take $B^{\alpha\beta} =\Sigma_{\gamma\beta}{}^\alpha = {\overline \Sigma}^\gamma{}_\beta{}^\alpha=0 $ and $A_\beta{}^\alpha = -\im \cS_\beta{}^\alpha$. Considering $d\bar\omega_Q$ equation in the same way, one finds also ${\overline A}_{\alpha}{}^\beta = \im \cS_\alpha{}^\beta$.

Then one should consider $d\omega_T$ equation \p{2NMCbos}. The $d\theta\times d\bar\theta$ terms in this equation read
\bea\label{N4dTeq2}
2\im \cS_\beta{}^\alpha \big( d_1\theta^\gamma d_2\bar\theta_\gamma + d_1\bar\theta_\gamma d_2\theta^\gamma  \big) = 2\im \big( d_1\theta^\alpha d_2\bar\theta_\gamma + d_1\bar\theta_\gamma d_2\theta^\alpha   \big)\cS_\beta{}^\gamma + \nn \\ + 2\im \big( d_1\theta^\gamma d_2\bar\theta_\beta + d_1\bar\theta_\beta d_2\theta^\gamma  \big)\cS_\gamma{}^\alpha-2\im \delta_\beta^\alpha \big(  d_1\theta^\nu d_2\bar\theta_\mu + d_1\bar\theta_\mu d_2\theta^\nu  \big)\cS_\nu{}^\mu.
\eea
These terms cancel, but to prove this it is necessary to take into account that $\alpha,\beta=1,2$, and these indices can be raised and lowered using the antisymmetric $\epsilon_{\alpha\beta}$, $\epsilon^{\beta\gamma}$ tensors. Then, if $M_\beta{}^\gamma=2\im \big( d_1\theta^\gamma d_2\bar\theta_\beta + d_1\bar\theta_\beta d_2\theta^\gamma  \big)$, one notes that
\bea\label{N4dTeq2smpl}
M_\gamma{}^\alpha \cS_\beta{}^\gamma + M_\beta{}^\gamma\cS_\gamma{}^\alpha - \delta_\beta^\alpha M_\mu{}^\nu \cS_\nu{}^\mu = \nn \\
=\big( M_\gamma{}^\alpha \cS_\beta{}^\gamma - M_{\gamma\beta}\cS^{\alpha\gamma} \big) +M_{\gamma\beta}\cS^{\alpha\gamma}+ M_\beta{}^\gamma\cS_\gamma{}^\alpha - \delta_\beta^\alpha M_\mu{}^\nu \cS_\nu{}^\mu = \nn \\
=M_{\gamma\beta}\cS^{\alpha\gamma}+ M_\beta{}^\gamma\cS_\gamma{}^\alpha = \big( -M_{\beta\gamma}+M_{\gamma\beta}  \big)\cS^{\alpha\gamma} = \cS_\beta{}^\alpha M_\mu{}^\mu,
\eea
which cancels the left hand side of \p{N4dTeq2}.
The $\triangle\tau\times d\theta$ and $\triangle\tau\times d\bar\theta$ terms in $d\omega_T$ equation read
\bea\label{N4dTeq3}
\triangle_1\tau d_2\theta^\gamma - \triangle_2\tau d_1\theta^\gamma: \;\; D_\gamma \cS_\beta{}^\alpha = 2\delta_\gamma^\alpha \bar\lambda_\beta -\delta_\beta^\alpha \bar\lambda_\gamma, \nn \\
\triangle_1\tau d_2\bar\theta_\gamma - \triangle_2\tau d_1\bar\theta_\gamma: \;\; \bD^\gamma \cS_\beta{}^\alpha = -2\delta_\beta^\gamma \lambda^\alpha +\delta_\beta^\alpha \lambda^\gamma.
\eea
Therefore, superfields $\cS^{\alpha\beta}$ satisfy the set of constraints of $\cN{=}4$, $d=1$ vector multiplet
\be\label{N4vect1}
D^{(\gamma}\cS^{\alpha\beta)} =0,\;\; \bD^{(\gamma}\cS^{\alpha\beta)} =0,\;\; \lambda^\alpha = \frac{1}{3}\bD^\gamma\cS_\gamma{}^\alpha, \;\; \bar\lambda_\alpha = - \frac{1}{3}D_\gamma\cS_\alpha{}^\gamma.
\ee

Analyzing the $d\omega_D$ equation, one quickly finds that the $d\theta\times d\bar\theta$ terms cancel, and others imply $\chi_\alpha = \im \bar\lambda_\alpha$, $\bar\chi_\alpha = -\im \lambda_\alpha$.

The $d\theta\times d\bar\theta$ terms also cancel from $d\omega_S$ equation, and the remaining $\triangle_1\tau d_2\theta^\alpha -\triangle_2\tau d_1\theta^\alpha$  term implies that
\be\label{N4dSeq2}
-\frac{1}{12}\big[ D_\mu, \bD_\nu \big]\cS^{\mu\nu} = -C - \frac{1}{2}\cS_{\mu\nu}\cS^{\mu\nu}.
\ee
To obtain this, one should use the vector multiplet conditions \p{N4vect1} to find
\bea\label{N4vectcons1}
&D_\beta\bD^\gamma\cS_\gamma{}^\alpha = -3\im \dot{\cS}_\beta{}^\alpha - \frac{1}{2}\delta_\beta^\alpha D_\mu\bD_\nu\cS^{\mu\nu}, \;\; \bD^\alpha D_\gamma\cS_\beta{}^\gamma = -3\im \dot{\cS}_\beta{}^\alpha + \frac{1}{2}\delta_\beta^\alpha D_\mu\bD_\nu\cS^{\mu\nu},&\nn \\
&D_\mu D_\nu \cS_\alpha{}^\beta = \bD^\mu\bD^\nu \cS_\alpha{}^\beta =0.&
\eea
Using this, it is easy to prove that $d\omega_K$ equation does not lead to any new conditions. Finally, all the forms are written in terms of $\cS_\alpha{}^\beta$ and its derivatives \p{N4formsfin}.

\end{document}